\input harvmac
\input epsf
\font\tay=eusb10
\font\twelverm=cmr10 scaled \magstep1
\font\tentit=cmmib10

\font\seventit=cmmib7
\font\fivetit=cmmib5
\newfam\titfam
\textfont\titfam=\tentit
\scriptfont\titfam=\seventit
\scriptscriptfont\titfam=\fivetit
\def\tit{\fam\titfam\tentit}
\def\CA{{\cal A}}  \def\CC{{\cal C}} \def\CD{{\cal D}}
 \def\CF{{\cal F}} \def\CG{{\cal G}} \def\CH{{\cal H}}
  \def\CK{{\cal K}} \def\CL{{\cal L}}
\def\CM{{\cal M}}  \def\CO{{\cal O}} \def\CP{{\cal P}}
  \def\CS{{\cal S}} \def\CT{{\cal T}}
 \def\CV{{\cal V}}  
 \def\CZ{{\cal Z}}
\def\rvec{{\bf \vec r}}
\def\kvec{{\bf \vec k}}

\def\rvec{{\bf \vec r}}
\def\rvecR{{\bf \vec r}_{\hbox{\fivebf R}}}
\def\bR{b_{\hbox{\fivebf R}}}
\def\kvec{{\bf \vec k}}

\def\qvec{{\bf \vec q}}

\def\Tay{{\hbox{\tay T}}}
\def\RR{\relax{\rm I\kern-.18em R}}
\def\Rr{\relax{\ninerm I\kern-.22em \ninerm R}}
\def\setminusp{\hbox{$/ \kern -3pt {}_p$}}
\def\ssetminusp{\hbox{$\scriptstyle / \kern -2pt {}_p$}}
%

%

\def\npr{\hbox{\twelverm :}}
\def\ii{{\rm i}}
\def\sprod{\mathop{\Pi}}
\def\sprodp{\mathop{\Pi'}}
\def\ssum{\mathop{\Sigma}}
\def\sssum{\mathop{{\scriptstyle\Sigma}}}
\def\blangle{\bigl\langle}
\def\brangle{\bigr\rangle}

\def\RR{\relax{\rm I\kern-.18em R}}
\def\Rr{\relax{\ninerm I\kern-.22em \ninerm R}}
\def\setminusp{\hbox{$/ \kern -3pt {}_p$}}
\def\ssetminusp{\hbox{$\scriptstyle / \kern -2pt {}_p$}}
\def\lesssim{\hbox{\raise.4ex \hbox{$<$} \kern-1.1em \lower.8ex \hbox{$\sim$}}}

\Title{\vbox{\hsize=3.truecm \hbox{}}}
{
\vbox{
\vskip -2.5truecm
\centerline{STATISTICAL MECHANICS}
\vskip .5truecm
\centerline{of}
\vskip .5truecm
\centerline{SELF-AVOIDING CRUMPLED MANIFOLDS}
\vskip 1.5truecm
\centerline{Part II}
\vskip .5truecm
}}
\vskip -.5cm
\centerline{{Bertrand Duplantier}\footnote{$^{\star}$}{e-mail: bertrand@spht.saclay.cea.fr}}
\bigskip\centerline{Service de Physique Th\'eorique\footnote{$^{\star \star}$}{
Laboratoire de la Direction des Sciences de la Mati\`ere du Commissariat \`a l'\'Energie Atomique, URA CNRS 2306.}}
\centerline{CEA/ Saclay}
\centerline{F-91191 Gif-sur-Yvette, France}

\vskip 1.truecm
\centerline{\bf Abstract{{}\footnote{$^{\dag}$}{Published in {\it Statistical Mechanics
of Membranes and Surfaces}, Second Edition, D. R. Nelson, T. Piran,
and S. Weinberg eds., World Scientific, Singapore (2004).}}}{
\ninerm
\textfont0=\ninerm
\font\ninemit=cmmi9
\font\sevenmit=cmmi7
\textfont1=\ninemit
\scriptfont0=\sevenrm
\scriptfont1=\sevenmit
\baselineskip=11pt
\bigskip
We consider a model of a $D$-dimensional tethered manifold interacting
by excluded volume in \Rr${}^d$ with a single point. Use of
intrinsic distance geometry provides a rigorous definition of the
analytic continuation of the perturbative expansion for arbitrary $D$,
\ $0\!<\!D\!<\!2$. Its one-loop renormalizability is first established by direct resummation. A renormalization operation
{\ninebf R} is then described, which ensures renormalizability to all orders. 
The similar question of the renormalizability of the self-avoiding manifold (SAM) Edwards model is
then considered, first at one-loop, then to all orders. 
We describe a short-distance multi-local operator product expansion,
which extends methods of local field theories to a large class of models with
non-local singular interactions. It vindicates the direct renormalization method used earlier in part {\bf I} of these lectures, 
as well as the corresponding 
scaling laws.}
\Date{}
\vskip .3in
\hfuzz 1.pt

\newsec{INTERACTING MANIFOLD RENORMALIZATION: A BRIEF HISTORY}
As can be seen in the set of lectures in this volume, which presents an extended version of \ref\Jerus{{\sl
Statistical Mechanics of Membranes and
Surfaces}, Proceedings of the Fifth Jerusalem Winter School for Theoretical
Physics (1987), D. R. Nelson, T. Piran, and S. Weinberg Eds., World Scientific,
Singapore (1989).}, the
statistical mechanics of random surfaces and membranes,
or more generally of extended objects, poses fundamental problems. The study of {\it polymerized}
membranes, which are generalizations of linear polymers 
\nref\SirSam{S. F. Edwards, Proc. Phys. Soc. Lond. {\bf 85} (1965) 613.}
\nref\desClJan{J. des Cloizeaux and G. Jannink,
{\sl Polymers in Solution, their Modelling and Structure}, Clarendon
Press, Oxford (1990).}
\nref\Wenetal{X. Wen, 
C.W. Garland, T. Hwa, M. Kardar,
E. Kokufuta, Y. Li, M. Orkisz, and T. Tanaka,
Nature {\bf 355} (1992) 426.}
\nref\Spector{M.S. Spector, E. Naranjo, S. Chiruvolu, and  J.A. Zasadzinski, Phys. Rev. Lett. {\bf 73} (1994) 2867.}
\nref\Spect{C.F. Schmidt, K. Svoboda, N. Lei, I.B. Petsche, L.E. Berman, C. Safinya, and G. S. Grest,
Science {\bf 259} (1993) 952.}
\nref\Stupp{S.I. Stupp, S. Son, H.C. Lin, and L.S. Li,  Science {\bf 259} (1993) 59.}
\nref\Reh{H. Rehage,  B. Achenbach, and A. Kaplan, Ber. Bunsenges. Phys. Chem. {\bf 101} (1997) 1683.}
\nref\KKN{Y. Kantor, M. Kardar, and D. R. Nelson, Phys. Rev. Lett. {\bf 57} (1986) 791; Phys. Rev. {\bf A 35} (1987) 3056.}
\nref\GK{See, {\it e.g.}, the lectures by G. Gompper and D.M. Kroll in this volume, and the references therein.}
\hskip-0.85cm [\xref\SirSam ,\xref\desClJan ] to two-dimensionally connected networks,
is emphasized,
with a number of possible experimental
realizations [\xref\Wenetal ,\xref\Spector ,\xref\Spect ,\xref\Stupp ,\xref\Reh ], or numerical simulations 
[\xref\KKN ,\xref\GK ]. From a theoretical point
of view, a clear challenge in the late eighties was to understand
self-avoidance (SA) effects in membranes.
\nref\KN{M. Kardar and D. R. Nelson, Phys. Rev. Lett. {\bf 58}
(1987) 1289, 2280(E); Phys. Rev. {\bf A 38}
(1988) 966.}
\nref\ArLub{J. A. Aronowitz and T. C. Lubensky, Europhys. Lett. {\bf 4}
(1987) 395.}\par

 The model proposed\foot{R.C.  Ball was actually the first to propose, while a postdoc in Saclay in 1981,
the  extension of the Edwards model to $D$-manifolds, with the aim, at that time,
to better understand polymers! (unpublished).} in [\xref\KN ,\xref\ArLub ]
aimed to incorporate the advances made in polymer theory
by renormalization group (RG) methods into the field of polymerized, or
tethered, membranes.  As we saw in part {\bf I} of these lectures,
these extended objects, {\it a priori} two-dimensional in nature,
are generalized for theoretical purposes to intrinsically
{\it $D$--dimensional manifolds} with internal points $x\in \RR^D$,
embedded in external $d$-dimensional space with position
vector $\rvec (x)\in \RR^d$.
The associated continuum Hamiltonian $\CH$ generalizes that of
Edwards for polymers \SirSam :
\eqn\Edwards{
\beta{\cal H}={1\over 2}\int d^Dx\,\Big(\nabla_x\rvec (x)\Big)^2+
{b\over 2}\int d^Dx\int d^Dx'\ \delta^{d}\big(\rvec (x)-\rvec (x')\big)
\ ,}
\nref\BDbis{B. Duplantier, Phys. Rev. Lett. {\bf 58} (1987) 2733.}
\hskip -3pt with an elastic Gaussian term and a self-avoidance
two-body $\delta$-potential with interaction parameter $b>0$. For
$0<D<2$, the Gaussian manifold ($b=0$) is {\it crumpled} with a Gaussian size exponent
\eqn\sizeo{{\nu}_0={2-D\over 2},} and a finite
Hausdorff dimension
\eqn\Haus{d_H=D/\nu_0=2D/(2-D);} the finiteness of the upper critical dimension
$d^\star = 2d_H$ for the SA-interaction
allows an $\varepsilon$-expansion about
$d^\star$ [\xref\KN --\nobreak\xref\BDbis ]:
\eqn\eps{\varepsilon=4D-2\nu_0 d}
performed via the direct renormalization method
adapted from that of des Cloizeaux in polymer theory
\ref\desCloiz{J. des Cloizeaux, J. Phys. France {\bf 42} (1981) 635.}, as we explained in part {\bf I}.

Only the polymer case,
with an {\it integer} internal dimension $D=1$, can be
mapped,
following de Gennes \ref\DeGe{P.G. de Gennes, Phys. Lett.
{\bf A 38} (1972) 339.},
onto a standard field theory, namely a $({\bf \Phi^2(\rvec)})^2$ theory
for an $n$-component field ${\bf \Phi}(\rvec)$ in external $d$-dimensional space, with $n\to 0$ components. This
is instrumental in showing that the direct renormalization
method for polymers is mathematically sound
\ref\BenMa{M. Benhamou and G. Mahoux, J. Phys. France {\bf 47} (1986) 559.},
and equivalent to rigorous renormalization schemes in standard
local field theory, such as the
Bogoliubov--Parasiuk--Hepp--Zimmermann (BPHZ) construction
\ref\BPHZ{N. N. Bogoliubov and O. S. Parasiuk, Acta Math. {\bf 97} (1957) 227;
\hfill\break\noindent
K. Hepp, Commun. Math. Phys. {\bf 2} (1966) 301;
\hfill\break\noindent
W. Zimmermann, Commun. Math. Phys. {\bf 15} (1969) 208.  }.
For manifold theory, we have to deal with {\it non-integer} internal
dimensions $D$, $D\ne 1$, and no such mapping exists.
Therefore, two outstanding problems remained in the theory
of interacting manifolds: (a) the mathematical meaning of a
{\it continuous} internal dimension $D$;
(b) the actual {\it renormalizability} of the perturbative
expansion of a manifold model like \Edwards ,
implying the scaling behavior expected on physical grounds.
\def\legend{(a) A $D$-manifold interacting with an impurity located at point {\bf 0} in $\RR^d$; (b) interaction with an 
Euclidean hyperplane of dimension $D'$ in $\RR^{d'}$, with $d'=d+D'$.}
\midinsert
\medskip
\centerline{\epsfxsize=12.truecm\epsfbox{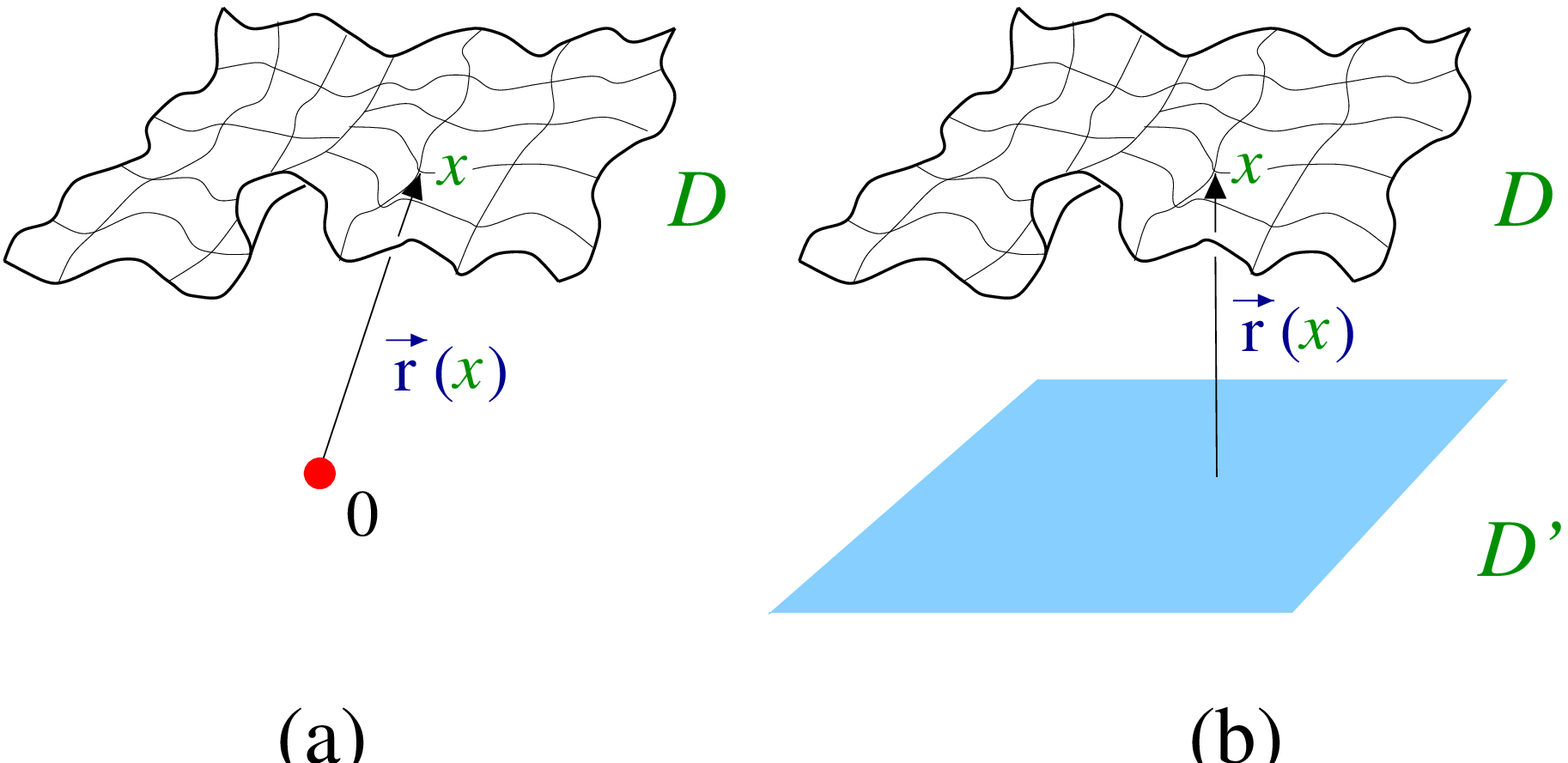}}
\medskip
\noindent
{\bf Fig. 1:\ }
{\sl \legend}
\endinsert
\nfig\fOne{\legend}
In
\ref\BD{B. Duplantier, Phys. Rev. Lett. {\bf 62} (1989) 2337.},
 a simpler model was proposed,
of a crumpled manifold interacting by excluded volume with
a fixed Euclidean subspace of $\RR^d$  \ref\rLasLip{See also M. L\"assig and R. Lipowsky, Phys. Rev. Lett. {\bf 70}
(1993) 1131.}.
The simplified model Hamiltonian introduced there reads:
\eqn\hami{
\beta{\cal H}={1\over 2}\int d^Dx\,\Big(\nabla_x \rvec (x)
\Big)^2+ {b}\int d^Dx\ \delta^{d}\big(\rvec (x)\big)
\ ,}
with a pointwise interaction of the
Gaussian manifold with an impurity located at the origin (Fig. 1a).
Note that this Hamiltonian also represents
interactions of a fluctuating (possibly directed) manifold
with a nonfluctuating $D'$- Euclidean hyperplane of $\RR^{d+D'}$,
$\rvec $ then standing for the coordinates transverse to this subspace (Fig. 1b).
The excluded volume case ($b>0$) parallels
that of the Edwards model \Edwards\ for SA-manifolds,
while an attractive interaction ($b<0$) is also possible, describing
pinning phenomena.
The (na{i}ve) dimensions of $\rvec$ and $b$ are respectively
$[\rvec ]=[x^\nu]$ with a Gaussian size exponent
\eqn\sizee{\nu\equiv (2-D)/2,} and
$[b]=[x^{-\varepsilon}]$ with
\eqn\epsi{\varepsilon\equiv D-\nu d.}
For fixed $D$ and $\nu$, the parameter $d$ (or equivalently $\varepsilon$)
controls
the relevance of the interaction, with the exclusion of a point
only effective for $d\le d^\star=D/\nu$. Note that in this model the size exponent $\nu$ is 
not modified by the local interaction and stays equal to its Gaussian value \sizee, whereas 
the correlation functions obey (non-Gaussian) universal scaling laws. 

For $D=1$, the model is exactly solvable \BD. For $D\ne 1$, the direct
resummation of leading divergences of the perturbation series is possible for model \hami\ and
indeed validates {\it one-loop} renormalization [\xref\BD].
This result was also extended to the Edwards model \Edwards\ itself
\ref\DHK{B. Duplantier, T. Hwa, and M. Kardar, Phys. Rev. Lett. {\bf 64}
(1990) 2022.}.

A study to all orders 
of the interaction model \hami\ was later performed in
\ref\DDGone{F. David, B. Duplantier, and E. Guitter,
{Phys. Rev. Lett.} {\bf 70} (1993) 2205, hep-th/9212102.},
\ref\DDGtwo{F. David, B. Duplantier, and E. Guitter, {Nucl. Phys.} {\bf B 394} (1993) 555,
cond-mat/9211038.}.
A mathematical
construction of the $D$-dimensional internal measure $d^Dx$
via distance geometry within the elastic manifold was given,
with expressions for manifold Feynman
integrals which generalize the $\alpha$-parameter representation
of field theory. In the case 
of the manifold model of [\xref\BD], the essential
properties which make it {\it renormalizable to all orders}
by a renormalization of the coupling constant were established. This led to a direct
 construction of a renormalization operation, generalizing the BPHZ construction to manifolds  
 (see also \ref\CPK{M. Cassandro and P.K. Mitter, Nucl. Phys. {\bf B 422} (1994) 634.} 
 \ref\PKS{P.K. Mitter and B. Scoppola, Commun. Math. Phys. {\bf 209} (2000) 207.}.)\par
 Later, the full Edwards model of self-avoiding manifolds \Edwards\ was studied by the same methods,
 and its renormalizability established to all orders \ref\DDGthree{F. David, B. Duplantier, and E. Guitter,
 {Phys. Rev. Lett.}
 {\bf 72} (1994) 311, cond-mat/9307059.},
 \ref\DDGfor{F. David, B. Duplantier, and E. Guitter, {\it Renormalization Theory for
 Self-Avoiding Polymerized Membranes}, cond-mat/9701136.}. Effective calculations to second order in $\varepsilon$
 (``two-loop'' order) were performed in
 \ref\DWone{F. David and K. Wiese, Phys. Rev. Lett. {\bf 76} (1996) 4564;
 {Nucl. Phys.} {\bf B 487} (1997) 529, cond-mat/9608022.}. The large order behavior of the
 Edwards model \Edwards\ was finally studied in \ref\DWtwo{F. David and K. Wiese, {Nucl. Phys.} {\bf B 535}
 (1998) 555, cond-mat/9807160.}.

The aim of part {\bf II} of these notes is to review some of these developments.

\newsec{MANIFOLD MODEL WITH LOCAL $\delta$ INTERACTION}
\subsec{\bf Perturbative expansion}
In this chapter, we study the statistical mechanics of the simplified model Hamiltonian \hami.
The model is described by its (connected) partition function
\eqn\PF{\CZ=\CV^{-1}\int {\cal D}[\rvec]\exp(-\beta{\cal H})} (here $\CV$
is the internal volume of the manifold)
and, for instance, by its one-point vertex function
\eqn\PFO{\CZ^{(0)}(\kvec )/\CZ=
\int d^Dx_0\ \langle e^{i\kvec\cdot\rvec(x_0)} \rangle,} where the
(connected) average
$\langle \cdots \rangle$ is performed with \hami:
\eqn\ZO{\CZ^{(0)}(\kvec )=\CV^{-1}\int {\cal D}[\rvec]\exp(-\beta{\cal H})\int d^Dx_0\ e^{i\kvec\cdot\rvec(x_0)}.}
These functions are all formally defined via their perturbative expansions
in the coupling constant $b$:
\eqn\PE{\CZ=\sum_{N=1}^{\infty}{(-b)^N\over N! }\,\CZ_N,} with a similar equation for
$\CZ^{(0)}$ with coefficients $\CZ_N^{(0)}$:
\eqn\PEO{\CZ^{(0)}(\kvec )=\sum_{N=1}^{\infty}{(-b)^N\over N! }\,\CZ_N^{(0)}(\kvec ).}

\def\legend{
Interaction points $x_i$; insertion point $x_0$ for the external momentum $\kvec $.}
\midinsert
\medskip
\centerline{\epsfxsize=5.truecm\epsfbox{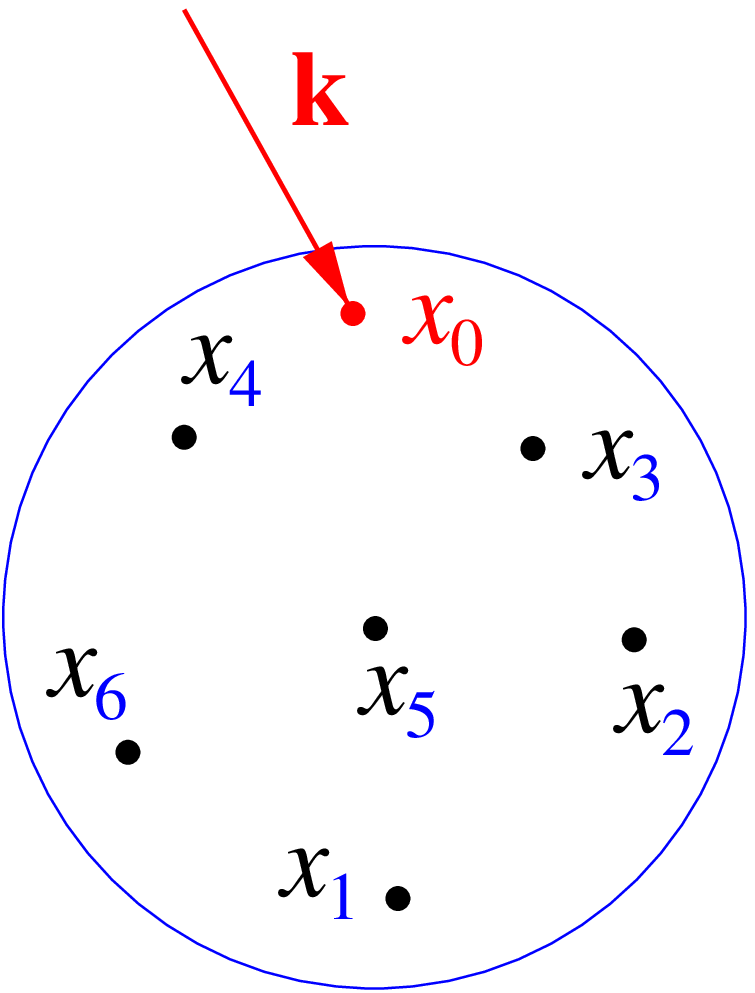}}
\medskip
\noindent
{\bf Fig. 2:\ }
{\sl \legend}
\endinsert
\nfig\fOne{\legend}
$\CZ_N$ has the path integral representation
\eqn\ZN{\CZ_N={1\over {\cal V}} \int d{\cal P}_0 \left[\int d^Dx\ \delta^{d}\big(\rvec (x)\big)\right]^N}
where the Gaussian path measure is
\eqn\dPO{d{\cal P}_0={\cal D}\rvec (x) \exp(-\beta {\cal H}_0)
}
with
\eqn\HO{\beta{{\cal H}_0}={1\over 2}\int d^Dx\,\Big(\nabla_x\rvec (x)\Big)^2.}
 There is no translational invariance in this theory, since the origin is selected by the presence of the impurity. 
 The measure $d{\cal P}_0$ thus {\it includes} integration over global translations of the manifold in $\RR^d$. The first term is 
 then simply 
$\CZ_1\equiv 1,$ so that
\eqn\Zb{\CZ=-b +{\cal O}(b^2).}
The term of order $N$, $\CZ_N$, is a Gaussian average
involving $N$ interaction points $x_i$ (Fig. 2):
\eqn\Zprod{\CZ_N={1\over {\cal V}} \int d{\cal P}_0 \int \prod_{i=1}^N
d^Dx_i\ \prod_{i=1}^N \delta^{d}\big(\rvec (x_i)\big).}
By Fourier transforming the distribution in $d$-space
$$\delta^{d}\big(\rvec (x)\big)=\int {d^d \kvec \over (2\pi)^d} \exp \left(i\kvec.\rvec\right),$$
one gets
\eqn\ZNI{{\CZ_N}={1\over {\cal V}}\int \prod_{i=1}^{N} d^Dx_i \int \prod_{i=1}^{N}{d^d \kvec_i\over (2\pi)^d} \int d{\cal P}_0
\exp\left[i\sum_{i=1}^{N}\kvec_i.\rvec_i\right].}
For a Gaussian manifold with weight \dPO\ \HO\ we have:
\eqn\gauss{\int d{\cal P}_0
\exp\left[i\sum_{i=1}^{N}\kvec_i.\rvec_i\right]=(2\pi)^d \delta^d\left(\sum_{i=1}^N\kvec_i\right)
\exp\left[-{1\over 2}\sum_{i,j=1}^{N}\kvec_i.\kvec_j G(x_i-x_j)\right].}
This Gaussian manifold average is expressed
solely in terms of the Green function
\eqn\green{G(x-y)=-{1\over 2} A_D|x\nobreak -\nobreak y|^{2\nu},}
solution\foot{In part {\bf I} we used the notation $G(x-y)\equiv A_D |x-y|^{2\nu}$ for the (positive) solution of the slightly different 
equation $\Delta_x G(x-y)=2\delta^D (x-y),$ while hereafter in {\bf II} we shall use the proper Newton-Coulomb potential \green,  
 in view of the underlying electrostatic representation.} of 
\eqn\poisson{-\Delta_x G(x-y)=\delta^D (x-y),}
with $2\nu=2-D$, and $A_D$ a normalization:
\eqn\norm{A_D=\left[S_D(2-D)/2\right]^{-1}=\left[S_D \nu\right]^{-1},}
where $S_D$ is the area of the unit sphere in $D$ dimensions
\eqn\SD{S_D={2\pi^{D/2}\over \Gamma(D)}.}
In the following, it is important to preserve the condition
$0<\nu <1$ ({\it i.e.}, $0<D<2$), corresponding to
the actual case of a crumpled manifold, where $(-G)$ is
positive and ultraviolet (UV) finite.\par

Performing finally the Gaussian integral over the $N-1$ independent real
variables $\kvec_i,\ (i=1,\cdots, N-1)$ yields [\xref\BD]:
\eqn\ZN{
\CZ_N\ ={\CV}^{-1}\,{(2\pi)^{-(N-1)d/2}}\, \int\prod_{i=1}^N d^Dx_i
\,\left(\det\left[ \Pi_{ij}\right]_{\scriptscriptstyle 1\le i,j\le  N-1 }
\right)^{-{d\over 2}},
}
where the matrix $[\Pi_{ij}]$ is simply defined as
\eqn\PI{\Pi_{ij}\,\equiv\,G({x_i-x_j})-G({x_i-x_N})-G({x_j-x_N}),}
with respect to the
reference point $x_N$, the permutation symmetry between the $N$ points being
restored in the determinant.\par

The integral representation of $\CZ_N^{(0)}$ is obtained from
that of $\CZ_N$ by multiplying the integrand in \ZN\ by
$\exp (-{1\over 2}\kvec^2 \Delta^{(0)} )$ with :
\eqn\forvertex{\Delta^{(0)}\equiv {\det[\Pi_{ij}]_{\scriptscriptstyle
0\le i,j \le N-1} \over
\det[\Pi_{ij}]_{\scriptscriptstyle 1\le i,j\le N-1} } \ ,}
and integrating over one more position, $x_0$, (Fig. 2):
\eqn\ZNO{
\CZ_N^{(0)}(\kvec)\ ={\CV}^{-1}\,{(2\pi)^{-(N-1)d/2}}\, \int\prod_{i=0}^N d^Dx_i\,\exp \left(-{1\over 2}\kvec^2 \Delta^{(0)} \right)
\,\left(\det\left[ \Pi_{ij}\right]_{\scriptscriptstyle 1\le i,j\le  N-1 }
\right)^{-{d\over 2}}.
}
Notice that the first order term ($N=1$) specializes to:
\eqn\ZNOone{
\CZ_1^{(0)}(\kvec)\ ={\CV}^{-1} \int_{\CV\times \CV} d^Dx_0\, d^Dx_1 \exp \left(-{1\over 2}\kvec^2 \Pi_{01} \right)
.
}
The resulting expressions are quite similar to those for the Edwards manifold model [\xref\DHK].

\medskip
\subsec{\bf Second virial coefficient} 
In this section, we imagine the manifold to be of finite internal volume ${\cal V}=X^D$, and define two dimensionless 
interaction coefficients, the excluded volume parameter $z$, and the second virial coefficient $g$, as
\eqn\z{z=(2\pi A_D)^{-d/2} b X^{D-(2-D)d/2},}
\eqn\g{g=(2\pi A_D)^{-d/2}(-\CZ) X^{D-(2-D)d/2}.}
Because of \Zb\ , the perturbative expansion of the full interaction parameter $g$ starts as:
\eqn\gz{g=z+{\cal O}(z^2).}
More precisely we have:
\eqn\gzc{g=\sum_{N=1}^{\infty} (-1)^{N-1} z^N I_N}
where we have set
\eqn\INbis{{1\over N!} \CZ_N \equiv (2\pi A_D)^{-(N-1)d/2}X^{(N-1)\varepsilon}I_N}
in order to get rid of cumbersome factors. Now the dimensionless integral $I_N$ is
\eqn\IN{I_N={1\over N!} \int_{\cal V'} \prod_{i=1}^{N} d^D x_i \ (\det {\bf D})^{-d/2},}
with integrations over {\it rescaled} coordinates, in a unit internal volume ${\cal V'}=X^{-D} {\cal V}=1$;  ${\bf D}$ is the symmetric 
$(N-1) \times (N-1)$ matrix with elements ($1\leq i,j\leq N-1$)
\eqn\Dij{\eqalign{
D_{ii}&=|x_{iN}|^{2-D}\cr D_{ij}&={1\over 2}\left(|x_{iN}|^{2-D}+|x_{jN}|^{2-D}-|x_{ij}|^{2-D}\right),\cr}}
where we set $x_{ij}\equiv x_i-x_j.$
\medskip
\subsec{\bf Resummation of leading divergences}
In this section we analyse the leading divergence of each $I_N$ for $\varepsilon=D-(2-D)d/2 =D-\nu d > 0$. We have 
$I_1=1$, and 
\eqn\Itwo{I_2={1\over 2} \int_{{\cal V'}\times {\cal V'}} d^D x_1\ d^D x_2\ |x_{1}-x_{2}|^{-(2-D)d/2}.}
We are interested in evaluating the pole at $\varepsilon=0$. It is easily extracted as [\xref\BD]
\eqn\Itoeps{I_2\simeq {1\over 2} \int_{{\cal V'}} d^D x_1 \int_{0}^{1}S_D\ d y\ y^{-1+\varepsilon}={S_D\over 2\varepsilon},}
where\foot{Note that the precise value of the upper limit for $y$,  $y\, \lesssim\, 1$, is immaterial 
when evaluating the pole part.} $y=|x_1-x_2|$.

The structure of divergences of the generic term $I_N$ will be studied in detail in the next sections. 
They will be shown to be only {\it local} divergences, obtained by letting any interaction point subset coalesce. Here, 
the leading divergence is evaluated as follows. 

The determinant in \IN\ is symmetrical with respect to the $N$ points, so we can, for a given $i \in \{1,\cdots, N-1\},$ 
and without loss of generality, consider the ``Hepp sector'' $x_i\to x_N$, hence $ \rho \equiv |x_{iN}|\to 0$. We then 
have  $D_{ii}=|x_{iN}|^{2-D}$, while  
 for any other  $j \in \{1,\cdots, N-1\},$ $j \neq i$, 
$$D_{ij}\simeq {1\over 2} \left(|x_{iN}|^{2-D} -x_{iN}.\nabla|x_{jN}|^{2-D}+{\cal O}(\rho^2)\right).$$ Using 
 $\nu= (2-D)/2$ and the notation 
$\delta\equiv\min (\nu, 1-\nu)$, we can write the leading term of this equation, which depends on the position of $D$,  
$0<D<2,$ 
with respect to 1, as 
$$D_{ij}=  |x_{iN}|^{\nu} \times {\cal O}\left(\rho^{\delta}\right).$$ 
When expanding the determinant $\det {\bf D}$ with respect to column $i$ and line $i$, we encounter either the diagonal term 
$D_{ii}=|x_{iN}|^{2\nu}={\cal O}(\rho^{2\nu})$,  
or non diagonal terms of type $D_{ij}D_{ik}={\cal O}\left(\rho^{2\nu +2\delta}\right)$. 
Thus $D_{ii}$ dominates and we can write 
in the sector $x_i\to x_N$
\eqn\fac{\det {\bf D} \simeq  D_{ii}\times \det {\bf D}/i= |x_{iN}|^{2-D}\times \det {\bf D}/i,} where $\det {\bf D}/i$ is the reduced determinant 
of order 
$(N-2) \times (N-2)$, in which  
line $i$ and column $i$ have been removed, hence the point $i$ itself. By symmetry, in any other sector $x_i\to x_j$, we have similarly 
\eqn\facg{\det {\bf D} \simeq  |x_{ij}|^{2-D}\times \det {\bf D}/i.} 
Among the $N(N-1)/2$ possible pairs $(i,j)$ we define an arbitrary ordered set of $N-1$ pairs 
${\cal P}=\{(i_{\alpha}, j_{\alpha}), \alpha=1,\cdots,N-1\},$ such that the distances  
$|x_{i_{\alpha}}-x_{j_{\alpha}}|=y_{\alpha} \to 0$ define a {\it sector} $y_1\leq y_2\leq \cdots \leq y_{N-1}$. In this limit,  
applying the rule \facg\ successively from $\alpha=1$ to $N-1$ yields a determinant factorized as 
$$\det {\bf D}\simeq \prod_{\alpha=1}^{N-1} y_{\alpha}^{2-D}.$$ The contribution of the sector ${\cal P}$ to the integral $I_N$ is 
given by the iteration of \Itoeps:

\eqn\sector{\eqalign{{I_N}_{|{\cal P}}&\simeq {1\over N!}\prod_{\alpha=1}^{N-1}\left[S_D \int_0^{y_{\alpha+1}} dy_{\alpha}\ y_{\alpha}^{-1+\varepsilon}
\right]\cr &={1\over N!}{1\over (N-1)!}\left[{S_D\over \varepsilon}\right]^{N-1}.\cr}}
The number of distinct sectors of $N-1$ ordered pairs ${\cal P}$ chosen among $N$ points equals $N! (N-1)!/2^{N-1}$, whence the leading divergence 
of $I_N$:
\eqn\lead{{I_N} \simeq \left({S_D\over 2\varepsilon}\right)^{N-1}.}
At this order, the dimensionless excluded volume parameter $g$ \g\ thus reads
\eqn\gsum{\eqalign{g&=\sum_{N=1}^{\infty} (-1)^{N-1} z^N I_N\simeq \sum_{N=1}^{\infty} (-1)^{N-1} z^N 
\left({S_D\over 2\varepsilon}\right)^{N-1}\cr &={z\over {1+z{S_D\over 2\varepsilon}}}.\cr}} 

\subsec{\bf Comparison to one-loop renormalization}
The Taylor-Laurent expansion of parameter $g$ to first orders is obtained from \gzc\ and \Itoeps
\eqn\gtwo{g=z-z^2 I_2+\cdots=z-z^2 {S_D\over 2\varepsilon}+\cdots.}
It is associated with a Wilson function
\eqn\W{\eqalign{W(g,\varepsilon)&=X{\partial g\over \partial X}=\varepsilon z{\partial g\over \partial z}\cr
&=\varepsilon z-z^2S_D+\cdots=\varepsilon g-g^2 {S_D\over 2}+\cdots.\cr}}
The fixed point $g^*$ such that $W(g^*,\varepsilon)=0$ is $g^*={2\varepsilon/ S_D}$ and precisely corresponds
to the limit of \gsum
\eqn\glim{g(z\to +\infty) ={2\varepsilon\over S_D}=g^*.}
More interestingly, the (truncated) flow equation \W
\eqn\Wtwo{\eqalign{W(g,\varepsilon)=\varepsilon z{\partial g\over \partial z}
=\varepsilon g-g^2 {S_D\over 2}\cr},}
with boundary condition \gz, has precisely the solution $g={z/ ({1+z{S_D\over 2\varepsilon}})}.$ 
So we see that the resummation \gsum\  to all orders of leading divergences is exactly equivalent to the 
one-loop renormalization group equation, as displayed in \Wtwo. Thus the one-loop renormalizability 
of the manifold model has been directly established by direct resummation of the perturbation expansion [\xref\BD].

This is confirmed by consideration of the vertex function \PFO. The same evaluation  
of \ZNO\ gives, after successive contractions of pairs of points in the determinants in \forvertex, \ZNO, the leading divergence:
\eqn\ZNObis{
\CZ_N^{(0)}(\kvec)\ ={1\over \CV}\,{(2\pi A_D)^{-(N-1)d/2}} X^{(N-1)\varepsilon}\, \left({S_D\over 2\varepsilon}\right)^{N-1} N! 
\int_{\CV \times\CV} d^Dx_0\ 
d^Dx_1\,\exp \left\{-{1\over 2}\kvec^2 \Pi_{01} \right\}
}
with the matrix element $\Pi_{01}= -2 G(x_0-x_1)$. 
The (connected) vertex function \ZO, \PEO\ can thus be resummed at this order as 
\eqn\PEObis{\eqalign{\CZ^{(0)}(\kvec)&=\sum_{N=1}^{\infty}{(-b)^N\over N! }\,\CZ_N^{(0)}(\kvec)\cr
&={-b\over 1+z {S_D\over 2\varepsilon}}\ {\CV}^{-1}
\int_{\CV \times\CV} d^Dx_0\ d^Dx_1\,\exp \left\{-{1\over 2}\kvec^2 (-2 G(x_0-x_1)) \right\}
 .\cr }}
 Notice that, at {\it first order}, $\CZ^{(0)}$ is determined  from \ZNOone\ as
\eqn\PEOter{\eqalign{\CZ^{(0)}(\kvec)
&=-b \CZ^{(0)}_1(\kvec)+{\cal O}(b^2)\cr&={-b}\ {\CV}^{-1}
\int_{\CV \times\CV} d^Dx_0\ d^Dx_1\,\exp \left\{-{1\over 2}\kvec^2 (-2 G(x_0-x_1)) \right\}+{\cal O}(b^2);
 \cr }}
 therefore the resummation of leading divergences in \PEObis\ amounts exactly to replacing
 $$b\to {b\over 1+z {S_D\over 2\varepsilon}}$$
 in the first order correlation function \PEOter. Owing to \z, this is indeed equivalent
 to replacing the bare dimensionless interaction parameter
 $z$ by the renormalized one $g=z/(1+z {S_D \over 2\varepsilon})$, in complete agreement with \gsum\ above.
\medskip
\subsec{\bf Analytic continuation in ${\tit D}$ of the Euclidean measure}
Integrals like \ZN\ or \ZNO, written with Cartesian coordinates, are {\it a priori} meaningful only for
integer $D$. Up to now, we have only formally extended such integrals to non-integer dimensions.  
Actually, an analytic continuation in $D$ can be performed
by use of {\it distance geometry} \DDGtwo. The key idea is to substitute
for the internal Euclidean coordinates $x_i$ the set of all mutual (squared)
distances $a_{ij}=(x_i-x_j)^2$ (Fig. 3).
\def\legend{Passage from Euclidean coordinates $x_i$ to the complete set of squared distances $a_{ij}$.}
\midinsert
\medskip
\centerline{\epsfxsize=12.truecm\epsfbox{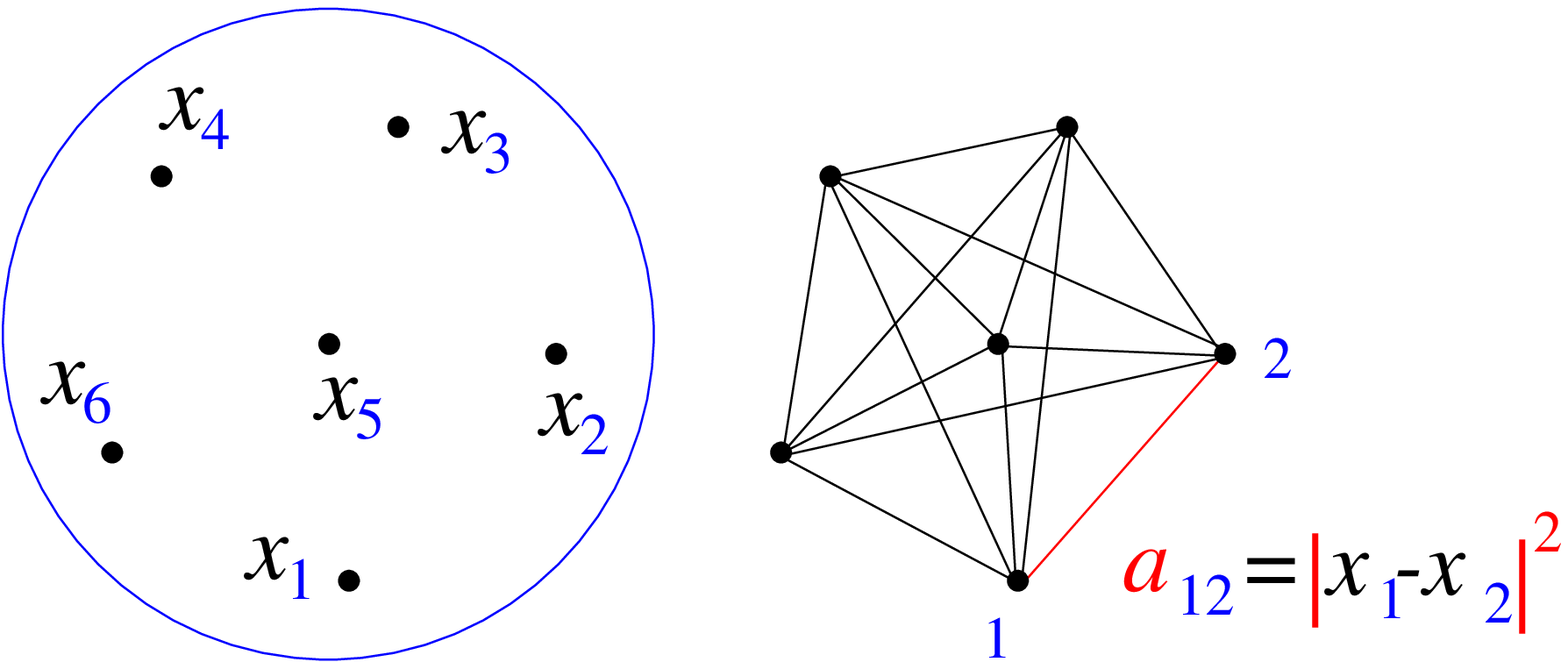}}
\medskip
\noindent
{\bf Fig. 3:\ }
{\sl \legend}
\endinsert
\nfig\fOne{\legend}
This is possible for integrands invariant
under the group of Euclidean motions (as in \ZN\ and \ZNO ). For
$N$ integration points, it also requires, {\it before} analytic continuation, $D$ to be large enough, {\it i.e.}, 
$D\ge N-1$, such that the $N-1$ relative vectors spanning these points
are linearly independent.

We define the graph $\CG$ as the set $\CG =\{1,\ldots ,N\}$
labelling the interaction points.
Vertices $i\in\CG$ will be remnants of the original Euclidean points
after analytic continuation, and index the squared distance matrix $[a_{ij}]$.
The change of variables $\{x_i\}_{i\in \CG}
\to a\equiv [a_{ij}]_{{i<j \hfill \atop i,j \in \CG}}$
reads explicitly \DDGtwo :
\eqn\inta{
{1\over \CV}\int_{\RR^D} \prod_{i\in \CG} d^Dx_i\,\cdots \ =\
\int_{{\cal A}_\CG}\, d\mu_\CG^{(D)}(a)\,\cdots
\ ,}
with the measure
\eqn\measaij{
d\mu_\CG^{(D)}(a)\equiv
\prod_{{i<j  \atop i,j \in \CG}}da_{ij}\
\Omega_N^{(D)}\, \Big(P_\CG(a) \Big)^{D-N\over 2}
\ ,}
where $N=|\CG|$, and
\eqn\Om{\Omega_N^{(D)}\equiv \prod_{K=0}^{N-2} {S_{D-K}\over 2^{K+1}}}
($S_D={2\pi^{D/2}\over \Gamma(D/2)}$
is as before the volume of the unit sphere in $\RR^D$), and
\eqn\CayMen{
P_\CG(a)\equiv {(-1)^N \over 2^{N-1}}\,
\left| \matrix{0&1&1&\ldots&1\cr 1&0&a_{12}&\ldots&a_{1N}\cr
1&a_{12}&0&\ldots&a_{2N}\cr \vdots&\vdots&\vdots&\ddots&\vdots\cr 1&a_{1N}&
a_{2N}&\ldots&0\cr } \right|\ . }
The factor
$\Omega_N^{(D)}$ \Om\ is the volume of the rotation group of the rigid
simplex spanning the points $x_i$. The ``Cayley-Menger determinant"
\ref\Blum{L. M. Blumenthal, {\sl Theory and Applications of Distance Geometry},
Clarendon Press, Oxford (1953).}
$P_\CG(a)$ is proportional to the squared Euclidean
volume of this simplex, a polynomial of degree $N-1$ in the $a_{ij}$.
The set $a$ of squared distances has to fulfill the triangular
inequalities and their generalizations:
$P_\CK(a)\ge 0$ for all subgraphs $\CK\subset \CG$,
which defines the domain of integration $\CA_\CG$ in \inta .

For real $D>|\CG|-2=N-2$, $d\mu_\CG^{(D)}(a)$ is a positive measure on $\CA_\CG$,
analytic in $D$.
It is remarkable that, as a distribution, it can be extended
to $0\le D\le |\CG|-2$ \DDGtwo . For integer $D\le |\CG|-2$,
although the change of
variables from $x_i$ to $a_{ij}$ no longer exists, Eq.\measaij\
still reconstructs the correct measure, concentrated on
$D$-dimensional submanifolds of $\RR^{N-1}$,
{\it i.e.}, $P_\CK=0$ if $D\le |\CK|-2$
\DDGtwo .
For example, when $D\to 1$ for $N=3$ vertices, we have,
denoting the distances $|ij|=\sqrt{a_{ij}}$:
\eqn\dmu{\eqalign{{d\mu^{(D\to 1)}_{\{1,2,3\}}(a)\over
d{\scriptstyle |12|}
d{\scriptstyle |13|}
d{\scriptstyle |23|}}
&=2\, \delta\big({\scriptstyle |12|+|23|-|13|}\big)
+2\, \delta\big({\scriptstyle |13|+|32|-|12|}\big)
+2\, \delta\big({\scriptstyle |21|+|13|-|23|}\big),}}
which indeed describes the 6 possibilities for nested intervals in $\RR$, with degeneracy factors 2
corresponding to the reversal of the 
orientation.
\medskip
Another nice feature of this formalism is that
{\it the interaction determinants in \ZN\ and \forvertex\ are also
  Cayley-Menger determinants!} We have indeed
\eqn\PiP{\det\left[\Pi_{ij}\right]_{1\le i,j \le N-1} = P_\CG(a^\nu)}
where $a^\nu\equiv[a_{ij}^\nu]_{{i<j\hfill \atop i,j \in \CG}}$ is obtained
by simply raising
each squared distance to the power $\nu$.
We arrive for \ZN\ and \ZNO\ at the representation of ``Feynman diagrams'' in
distance geometry:
\eqn\ZNa{\eqalign{
&\CZ_N=\int_{\CA_\CG}d\mu^{(D)}_\CG\, I_\CG\ ,\ \ \
I_{\CG} =\big(P_\CG(a^\nu)\big)^{ -{d\over 2}}  \cr
&\CZ_N^{(0)}(\kvec)=
\int_{\CA_{\CG \cup \{0\}}}d\mu^{(D)}_{\CG \cup \{0\}}
\, I_\CG^{(0)}(\kvec) \ ,\cr
&I_{\CG}^{(0)}(\kvec) =
I_\CG \ \exp \left(-{1\over 2}\kvec^2
{P_{\CG\cup\{0\}}(a^\nu)\over P_\CG(a^\nu)}
\right)\hfill ,\cr } }
which are $D$-dimensional extensions of the Schwinger $\alpha$-parameter
representation.
We now have to study the actual
convergence of these integrals and, possibly, their renormalization.
\medskip
\subsec{\bf Analysis of divergences}
Large distance infrared (IR) divergences occur for
manifolds of infinite size. One can keep a
finite size, preserve symmetries and
avoid boundary effects by choosing as a manifold
the $D$-dimensional sphere $\CS_D$ of radius $R$
in $\RR^{D+1}$. This amounts \DDGtwo\ in distance geometry
to substituting for $P_\CG(a)$ the ``spherical" polynomial
$P_\CG^{\CS}(a)\equiv  P_\CG(a)+{1\over R^2}\det(-{1\over 2}a)$,
the second term providing an IR cut-off, such that $a_{ij}\le 4R^2$.
In the following, this IR regularization
will simply be ignored when dealing with
short-distance properties, for which we can take $P_\CG^\CS\sim P_\CG$. This was also the case when evaluating
 leading divergences in the sections above.

The complete description of the possible set of divergences is then obtained from the
following theorem of distance geometry \Blum :\par
\smallskip

\noindent{\it Schoenberg's theorem}.
{\sl For $0<\nu<1$, the set
$a^\nu=[a^\nu_{ij}]_{{ i<j\hfill \atop i,j \in \CG}}$ can be realized as the
set of
squared distances
of a transformed simplex in $\RR^{N-1}$, whose volume
$P_\CG(a^\nu)$ is positive, and vanishes if and only if at least one of
the mutual
original distances itself vanishes}, $a_{ij}=0$.\par
\smallskip

\noindent This ensures that, as in field theory, the only source
of divergences in $I_\CG$ and $I_\CG^{(0)}$ is at {\it short distances}.
Whether these UV singularities are integrable or not will depend
on whether the external space dimension $d< d^\star
=D/\nu$ or $d>d^\star$.

\subsec{\bf Factorizations} The key to convergence and
renormalization is the following
short-distance {\it factorization} property of $P_\CG(a^\nu)$.
Let us consider a subgraph $\CP\subset\CG$, with at least two vertices,
in which we
distinguish an element, the {\it root} $p$ of $\CP$, and let us denote by
$\CG\setminusp\CP \equiv (\CG\setminus\CP )\cup \{p\}$ the subgraph
obtained by replacing in $\CG$ the whole
subset $\CP$ by its root $p$.
In the original Euclidean formulation, the analysis of short-distance
properties amounts to that of contractions of points $x_i$, labeled
by such a subset $\CP$, toward the point $x_p$, according to:
$x_i(\rho )=x_p+\rho (x_i-x_p)$ if $i\in \CP$,
where $\rho\to 0^{+}$ is the dilation factor, and
$x_i(\rho)=x_i$ if $i\notin \CP$.
This transformation has
an immediate resultant 
in terms of mutual distances:
$a_{ij}\ \to a_{ij}(\rho)$, depending on both $\CP$ and $p$.
Under this transformation, the interaction polynomial
$P_\CG(a^\nu)$ factorizes into \DDGtwo :
\eqn\factdet{\eqalign{
P_\CG(a^\nu (\rho)) & =
P_\CP(a^\nu(\rho ))\,
P_{\CG\setminusp \CP}(a^\nu) \cr & \quad \quad  \times
\left\{1+{\cal O}(\rho^{2\delta})\right\}\
\ .\cr}}
with $ \delta=\min(\nu, 1-\nu)>0 $ and
where, by homogeneity,
$P_\CP(a^\nu(\rho ))=\rho^{2\nu (|\CP |-1)}\, P_\CP(a^\nu)$.
\def\legend{Factorization property \factdet .}
\midinsert
\medskip
\centerline{\epsfbox{contraction.eps}}
\medskip
\noindent
{\bf Fig. 4:\ }
{\sl \legend}
\endinsert
\nfig\fOne{\legend}
\noindent The geometrical interpretation of \factdet\ is quite
simple:
the contribution of the set $\CG$ splits into that of
the contracting subgraph $\CP$ multiplied by that of the whole set $\CG$ where
$\CP$ has been replaced by its root $p$ (Fig. 4),
all correlation distances between these subsets being suppressed. The factorization property \factdet\ is the
generalization, to an arbitrary set $\CP$ of contracting points, of the factorization  encountered in \facg\ for the contraction
of a pair of points. This is simply, in this interacting manifold model,
the rigorous expression of an {\it operator
product expansion} \DDGtwo .

The factorization property \factdet\ does not hold for $\nu = 1$,
preventing a factorization of the measure \measaij\
$d\mu^{(D)}_{\CG }(a)$ itself.
Still, the integral of the measure, when applied to a factorized
integrand, does factorize as:
\eqn\factint{\int_{\CA_\CG}d\mu^{(D)}_\CG\cdots=\int_{\CA_\CP}
d\mu^{(D)}_\CP\cdots\int_{\CA_{(\CG\ssetminusp \CP)}}
d\mu^{(D)}_{(\CG\setminusp\CP)}\cdots \ .}
This fact, explicit for integer $D$ with a
readily factorized measure $\prod_i d^Dx_i$,
is preserved \DDGtwo\ by analytic
continuation only after integration over relative distances between the
two ``complementary" subsets $\CP$ and $\CG\setminusp \CP$.
\medskip
\subsec{\bf Renormalization}
A first consequence of factorizations \factdet\ and \factint\ is
the absolute convergence of $\CZ_N$ and $\CZ_N^{(0)}$ for $\varepsilon >0$.
Indeed, the superficial degree of divergence of $\CZ_N$ (in distance units)
is $(N-1)\varepsilon$, as can be read from \ZNa , already ensuring the
superficial convergence when $\varepsilon>0$.
The above factorizations ensure that the superficial degree of
divergence in $\CZ_N$ or $\CZ_N^{(0)}$ of any subgraph $\CP$ of $\CG$
is exactly that of $\CZ_{|\CP |} $ itself, {\it i.e.}, $(|\CP |-1)\varepsilon >0$.
By recursion, this ensures the absolute convergence of the manifold
Feynman integrals. A complete discussion has recourse to a generalized
notion of Hepp sectors and is given in \DDGtwo . In the proof,
it is convenient to first consider $D$ large enough
where $d\mu_{\CG}^{(D)}$ is a non-singular measure,
with a fixed $\nu$ considered as an independent variable $0<\nu<1$,
and to then continue to $D=2-2\nu$, $0<D<2$,
corresponding to the physical case.

When $\varepsilon =0$, the integrals giving $\CZ_N$ and $\CZ_N^{(0)}$
are (logarithmically) divergent.
Another consequence of Eqs. \factdet\ and \factint\ is thus 
the possibility
to devise a renormalization operation {\bf R},
as follows.
To each contracting rooted subgraph $(\CP,p)$ of $\CG$,
we associate a Taylor operator $\Tay_{(\CP,p)}$, performing
on interaction integrands the exact factorization corresponding to \factdet :
\eqn\TayI{\Tay_{(\CP,p)}I_\CG^{(0)}=I_{\CP}\,I_{\CG\setminusp \CP}^{(0)}
\ ,}
and similarly
$\Tay_{(\CP,p)}I_\CG=I_{\CP}\,I_{\CG\setminusp \CP}$.
As in standard field theory \BPHZ ,
the subtraction renormalization operator {\bf R}
is then organized in terms of forests \`a la Zimmermann.
In manifold theory, we define
a {\it rooted forest} as a set of rooted subgraphs $(\CP,p)$ such
that any two subgraphs are either disjoint or nested, {\it i.e.}, never
partially overlap. Each of these subgraphs in the forest
will be contracted toward its root under the action \TayI\ of the
corresponding Taylor operator.
When two subgraphs $\CP\subset \CP'$ are nested, the smallest one
is contracted first toward its root $p$, the root
$p'$ of $\CP'$ being itself attracted toward $p$
if $p'$ happened to be in $\CP$.
This hierarchical structure is anticipated by choosing the roots of the
forest as {\it compatible}: in the case described above, if $p'\in \CP$,
then $p'\equiv p$.
Finally, the renormalization operator is written
as a sum over all such compatibly rooted forests of $\CG$, denoted by
$\CF_\oplus$:
\eqn\Roper{
{\bf R}\ =\
\sum_{\CF_{\oplus}}\,W(\CF_{\oplus})\,
\Bigg[ \prod_{(\CP,p)\in \CF_{\oplus}}
\!\big( -\Tay_{(\CP,p)} \big)\Bigg]
\ .}
Here $W$ is a necessary combinatorial weight associated with the degeneracy
of compatible rootings, $W(\CF_{\oplus})\,
=\, \prod_{ {p\ {\rm root}}\atop {{\rm of}\,\CF_{\oplus}} }
1 /|\CP(p)| $ with $\CP(p)$ being the largest subgraph of the forest
$\CF_\oplus$ whose root is $p$.
An important property is that, with compatible roots, the Taylor operators
of a given forest now commute \DDGtwo.
The renormalized amplitudes are defined as
\eqn\ZMRen{
{\CZ^{\bf R}}_N^{(0)}(\kvec)\ \equiv \
\int_{\CA_{\CG \cup \{0\}}} d\mu^{(D)}_{\CG\cup\{0\}}\,{\bf R}\,[I_\CG^{(0)}(\kvec)]
\ .}
The same operation ${\bf R}$
acting on $I_\CG$ leads automatically by homogeneity to
$ {\bf R}\left[I_{\CG}\right]=0 $ for $|\CG|\ge 2$.
We state the essential result that now {\it the renormalized Feynman
integral} \ZMRen\ {\it is convergent}: ${\CZ^{{\bf R}}}_N^{(0)} < \infty$
for $\varepsilon=0$.
A complete proof of this renormalizability property  is given in \DDGtwo\
the analysis being
inspired from the direct proof by Berg\`ere and Lam of the
renormalizability in field theory of Feynman amplitudes
in the $\alpha$-representation
\ref\BergLam{M. C. Berg\`ere and Y.-M. P. Lam,
J. Math. Phys. {\bf 17} (1976) 1546.}.
\medskip
The physical interpretation of the renormalized amplitude \ZMRen\ and
of \Roper\ is simple.
Equations \factint\ and \TayI\ show that
the substitution for the bare amplitudes \ZNa\ of the renormalized ones
\ZMRen\ amounts to a reorganization to all orders of the original
perturbation series in $b$,
leading to the remarkable identity:
\eqn\RenExp{
\CZ^{(0)}(\kvec)\ =\ \sum_{N=1}^\infty \,{(-b_{\bf R})^N\over N!}\,
{\CZ^{\bf R}}_N^{(0)}(\kvec)
\ ,}
where the {\it renormalized} interaction parameter $b_{\bf R}$ is simply here (minus) the connected partition function
\eqn\br{b_{\bf R}\equiv -\CZ.}
This actually extends
to any vertex function, showing that the theory is
made perturbatively finite (at $\varepsilon = 0$)
by a full renormalization of the coupling constant $b$
into $-\CZ$ itself, in agreement with the definition of the second virial coefficient $g$ \g\ above.  From this result,
one establishes the existence to all orders of the Wilson function \W
$$W(g,\varepsilon)={X} {\partial g \over \partial X}{\big|_b},$$
describing the scaling properties of the interacting manifold
for $\varepsilon$ close to zero, and which has a {\it finite limit} up to $\varepsilon=0$ \DDGtwo .
For $\varepsilon >0$, an IR
fixed point at $b>0$ yields universal excluded volume exponents;
for $\varepsilon <0$, the associated UV fixed point at $b<0$ describes
a localization transition.

This demonstrated how to define an interacting manifold model with
continuous internal dimension, by use of distance geometry, as a natural
extension of the Schwinger representation for field theories. Furthermore,
in the case of a pointwise interaction, the
manifold model is indeed renormalizable to all orders.
The main ingredients are Schoenberg's theorem
of distance geometry, insuring that divergences occur only at short
distances for (finite) manifolds, and the short-distance factorization
of the generalized Feynman amplitudes.
This provided probably the
first example of a perturbative renormalization
established for extended geometrical objects \DDGtwo.
This opens the way to the renormalization theory of self-avoiding manifolds, 
which we now sketch.


\newsec{SELF-AVOIDING MANIFOLDS \& EDWARDS MODELS}
\subsec{\bf Introduction}
In this part, we concentrate on the renormalization theory of
the model of tethered self-avoiding manifolds (SAM) [\xref\KN,\xref\ArLub],
directly inspired by the Edwards model for polymers \SirSam:
\eqn\eEdwards{
{\raise.2ex\hbox{$\CH$}/\raise -.2ex\hbox{${\rm k}_{\rm B}T$}}
\ =\ {1\over 2}\,\int d^Dx\,\big(\nabla_x\rvec (x)\big)^2+
{b\over 2}\int d^Dx\int d^Dx'\ \delta^{d}\big(\rvec (x)-\rvec (x')\big)
\ ,}
with an elastic Gaussian term and a self-avoidance
two-body $\delta$-potential with excluded volume parameter $b>0$.
Notice that in contrast with the local $\delta$ interaction model \hami\ studied in $\S\, 2$, the interaction here is {\it non-local} in
``manifold space" $\RR^D$.

The finite upper critical dimension (u.c.d.) $d^\star$ for the SA
interaction exists only for manifolds with a continuous internal
dimension $0<D<2$. For $D\to 2$, $d^\star \to +\infty$.
Phantom manifolds ($b=0$) are {\it crumpled} with a finite
Hausdorff dimension $d_H=2D/(2-D)$, and $d^\star = 2d_H$.
The $\varepsilon$-expansion about $d^\star$  performed in [\xref\KN,\xref\ArLub,\xref\BDbis], and described 
in part {\bf I} above, was directly inspired by the des Cloizeaux
{\it direct renormalization} (DR) method in
polymer theory \desCloiz. But the issue of  the consistency of the DR method remained unanswered, since 
for  $D\ne 1$, model \eEdwards\ cannot be mapped onto a standard $({\bf \Phi}^2 (\rvec))^2$ local field
theory.

The question of  {\it boundary effects} in relation to the value of the
configuration exponent $\gamma$  also requires some study \BDbis.
It caused some confusion in earlier publications [\xref\KN,\xref\ArLub ,\xref\BDbis]. In part {\bf I} of these lectures, we
showed that a  finite self-avoiding patch embedded in an infinite Gaussian manifold has exponent $\gamma =1$ for any
$0<D<2,\ D\ne 1$. Here the cases of closed or open manifolds with free boundaries will be considered.

\subsec{\bf Renormalizability to first order}
The validity of RG methods and of scaling laws was first justified
 at leading order in $\varepsilon$ through explicit resummations in \DHK, in close analogy to the
 procedure described in $\S\, 2.3$ above for the $\delta$-interaction impurity model.
 We shall not repeat all the arguments here, but comment on some significant results.

 Let us  consider the spatial correlation function
 $\blangle [\rvec(x)-\rvec(0)]^2\brangle.$ For a Gaussian (infinite) manifold it equals
\eqn\RGauss{\blangle [\rvec(x)-\rvec(0)]^2\brangle_{0}=d\; [-2G(x)]=d\; A_D |x|^{2-D}=d{2\over S_D(2-D)} |x|^{2-D}.}
In the presence of self-avoidance, it is expected to scale as:
\eqn\cor{\blangle [\rvec(x)-\rvec(0)]^2\brangle\propto |x|^{2\nu},}
with a swelling exponent $\nu \geq \nu_0=(2-D)/2$ for $d\leq d^{*}$. It can be directly evaluated by
resummation of leading divergences \DHK:
\eqn\corresum{\blangle [\rvec(x)-\rvec(0)]^2\brangle =d{2\over S_D(2-D)} |x|^{2-D}
\left(1+{a\over \varepsilon} b_D |x|^{\varepsilon/2}\right)^{a_0/a},}
where $b_D$ is simply the bare interaction parameter $b$ conveniently dressed by coefficients
$$b_D= (2\pi A_D)^{-d/2}b=\left[4\pi/S_D(2-D)\right]^{-d/2}b,$$
and where $a_0$ and $a$ are two universal coefficients \DHK:
\eqn\aa{a_0={S_D^2\over D}{2-D\over 2},\;\;\; a=S_D^2\left(1+{1\over 2-D}{\Gamma^2({D/(2-D)})\over \Gamma ({2D/(2-D)})}\right),}

The scaling behavior \cor\ is then directly recovered  from \corresum\ in the
large distance or strong self-avoidance limit $b |x|^{\varepsilon/2}\to +\infty$, with
 a value of the swelling exponent $\nu$ at first order in $\varepsilon$:
\eqn\NU{\nu={2-D\over 2 }+{1\over 2}{a_0\over a}{\varepsilon\over 2},}
or explicitly:
\eqn\NUexp{\nu = {2-D\over 2 }\left\{1+{\varepsilon\over2}{1\over 2D}\left[1
+{1\over 2-D}{\Gamma^2({D/(2-D)})\over \Gamma ({2D/(2-D)})}\right]^{-1}\right\},}
in agreement with the result (3.24) of part {\bf I}.

Similarly, for a manifold of finite volume $\CV=X^D$, one defines
a dimensionless excluded volume parameter $z$, as in part {\bf I} of these lectures, by
\eqn\zsam{z=b_D X^{2D-(2-D)d/2}=(2\pi A_D)^{-d/2}b X^{\varepsilon/2}.}
One finds an effective size of the membrane:
\eqn\Rtwo{R^2=\blangle \left [\rvec(X)-\rvec(0)\right ]^2\brangle={\cal X}_0(z,\varepsilon)\ d{2\over S_D(2-D)} X^{2-D}}
where ${\cal X}_0(z,\varepsilon)$ is the {\it swelling factor} with respect to the Gaussian size 
 $\blangle [\rvec(X)-\rvec(0)]^2\brangle_0 $ \RGauss,
as introduced in part {\bf I}, Eq. (3.1). The direct
resummation of  leading divergences to all perturbative orders gives \DHK:
\eqn\swell{{\cal X}_0(z,\varepsilon)=\left(1+{a\over \varepsilon} b_D X^{\varepsilon/2}\right)^{a_0/a}
=\left(1+{a\over \varepsilon} z\right)^{a_0/a}.}
At first order in $z$, we recover
\eqn\swellone{{\cal X}_0(z,\varepsilon)=1+{a_0\over \varepsilon} z+ {\cal O}(z^2),}
which is the perturbative result (2.44) of part {\bf I}.

We also intoduced in part {\bf I}, Eqs. (3.7-9), the {\it dimensionless second virial coefficient} $g$
\eqn\gsam{g=-(2\pi R^2/d)^{-d/2} {\CZ_{2,c}\over \CZ_1^2},}
where  $\CZ_1$ and $\CZ_{2,c}$ are respectively the (connected) 1-manifold and 2-manifold partition functions.
The same direct resummation of leading divergences in perturbation theory  gives
for $g$
\eqn\gsamresum{g={z\over 1+{az/ \varepsilon}},}
with a first order expansion
\eqn\gsamone{g=z- z^2 {a/ \varepsilon}+\cdots,} in agreement with {\bf I.} Eqs. (3.16-17) [$a$ was noted as
${a'}_D$ there.]

 It is interesting to observe the following fact, key to a rigorous approach to renormalizablity to first order.
 The RG flow equations were obtained in part {\bf I.} Eqs. (3.4) (3.19) (3.22) from first order results
 (here {\bf II.} \swellone, \gsamone) for the scaling functions
\eqn\flow{\eqalign{W(g,\varepsilon)=X{\partial g\over \partial X}={\varepsilon\over 2} z{\partial g\over \partial z}
={\varepsilon\over 2} z-z^2 a+\cdots={\varepsilon\over 2} g-g^2 {a\over 2}+{\cal O}(g^2) \cr}}
\eqn\flowbis{X{\partial \over \partial X}\ln {\cal X}_0(z,\varepsilon)
={\varepsilon\over 2} z{\partial \over \partial z}\ln {\cal X}_0(z,\varepsilon)
={1\over 2}a_0\; z+\cdots={1\over 2}a_0\; g+{\cal O}(g^2).}
When truncated to this order, their solutions are exactly the resummed expressions \gsamresum\ and \swell.
Turning things around,
 the direct resummation of leading poles in $\varepsilon$ indeed establishes one-loop renormalizability [\xref\BD,\xref\DHK].

\medskip
\subsec{\bf Renormalizabilty to all orders}
We briefly describe below the formalism that allows to prove
the validity of the RG approach to self-avoiding manifolds,
as well as to a larger class of manifold models with non-local interactions.
(See \DDGtwo, \DDGfor, for further details.). This formalism is based on an operator product expansion involving
{\it multi-local singular operators}, which allows a systematic analysis of
the short-distance ultraviolet singularities of the Edwards model.
At the critical dimension $d^\star$, one can classify all of the relevant operators
and show that the model \eEdwards\ is {\it renormalizable to all orders} by
renormalizations (i) of the coupling $b,$  
and (ii) of the position field $\rvec$.
As a consequence, one establishes the validity of scaling laws for {\it infinite}
membranes, as well as the existence of finite size scaling laws for
{\it finite} membranes.
The latter result ensures the consistency of the DR approach.

A peculiar result, which distinguishes manifolds with non-integer $D$ from
open linear polymers with $D=1$, is the absence of {\it boundary} operator
multiplicative renormalization, leading to the general {\it hyperscaling relation} for the configuration exponent $\gamma$
\eqn\eHyper{\gamma=1-\nu d\, ,}
valid for finite SAM with $0<D<2$, $D\ne 1$. Note that this hyperscaling value is also
valid for {\it closed} linear polymers (see, {\it e.g.}, \ref\BDhype{B. Duplantier, 
{Nucl. Phys.} {\bf B 430} (1995) 489.}.) This result
is valid for closed or open manifolds with free boundaries, and has the same origin
as the result $\gamma=1$ obtained in part {\bf I} for a finite SA patch  embedded in an infinite manifold
(see {\bf I.} $\S\, 2.2.2$ and $\S\, 2.2.3.$)

\subsec {\bf Perturbation theory and dipole representation}
As in part {\bf I}, the partition function is defined  by the
functional integral:
\eqn\ePartFunc{
\CZ\ =\ \int \CD[\rvec(x)]\,\exp\left(
-{\raise.2ex\hbox{$\CH[\rvec]$}/\raise -.2ex\hbox{${\rm k}_{\rm B}T$}}
\right)
\ .}
It has a perturbative expansion in $b$, formally given by
expanding the exponential of the contact interaction
\eqn\eZPertExp{\eqalign{
\CZ\ &=\ \CZ_0\ \sum_{N=0}^\infty\,{(-b/2)^N\over N!}\,
\int\sprod_{i=1}^{2N}{d^Dx_i}\ 
\bigl\langle
\sprod_{a=1}^{N}\delta^{d}(\rvec(x_{2a})-\rvec(x_{2a-1}))
\bigr\rangle_0\cr
&\equiv\ \CZ_0\ \sum_{N=0}^\infty\,{(-b/2)^N\over N!}\, Z_N\ ,\cr
}}
where $\CZ_0$ is the partition function of the Gaussian manifold (hence $Z_0 \equiv 1$), 
and
$\langle\cdots\rangle_0$ denotes the average with respect to the
Gaussian manifold ($b=0$):
\eqn\evevo{
\langle(\cdots)\rangle_0\ =\ {1\over\CZ_0}\,
\int \CD[\rvec(x)]\,\exp\left(-{1\over 2}\int d^Dx\,
(\nabla_x\rvec(x))^2\right)(\cdots)
\ .
}

Physical observables are provided by average values of operators, which
must be invariant under global translations.
Using Fourier representation,
local operators can always be generated by the exponential operators
(or vertex operators), of the form
\eqn\eVertOp{
V_{\qvec}(z)\ =\ {\rm e}^{\ii\qvec\cdot\rvec(z)}
\ .
}
In perturbation theory the field $\rvec(x)$ will be treated as a massless free
field and the momenta $\qvec$ will appear as the
``charges" associated with the translations in $\RR^d$.
Translationally invariant operators are then provided by ``neutral" products
of such local operators,
\eqn\eNeutObs{
O_{\qvec_1,\cdots,\qvec_P}(z_1,\cdots,z_P)\ =\ \prod_{l=1}^{P}\,
V_{\qvec_l}(z_l)\ ,
\qquad \qquad \qvec_{\rm total}\ =\ \sum_{l=1}^{P}\,\qvec_l\ =\ {\vec{\bf 0}}
\ .
}
The perturbative expansion for these observables is simply
\eqn\eCorrFirst{\eqalign{
\bigl\langle \sprod_{l=1}^{P}{\rm e}^{\ii \qvec_l\cdot\rvec(z_l)}\bigr\rangle
\ &=\ {1\over \CZ}\,\sum_{N=0}^\infty\,{(-b/2)^N\over N!}\,
\int\sprod_{i=1}^{2N}{d^Dx_i}
\bigl\langle
\sprod_{l=1}^{P}{\rm e}^{\ii \qvec_l\cdot\rvec(z_l)}
\sprod_{a=1}^{N}\delta^{d}(\rvec(x_{2a})-\rvec(x_{2a-1}))
\bigr\rangle_0 \cr
&\equiv\ {1\over \CZ}\,\sum_{N=0}^\infty\,{(-b/2)^N\over N!}\, Z_N\left(\{\qvec_l\}\right)\ .\cr
}}
Each $\delta$ function in \eZPertExp\ and \eCorrFirst\
can itself be written in terms of two exponential operators as
\eqn\eExpRep{
\delta^{d}(\rvec(x_{2})-\rvec(x_{1}))\ =\
\int{d^d\kvec_{1}d^d\kvec_{2}\over (2\pi)^d}
\,\delta^d(\kvec_{1}+\kvec_{2})
\,{\rm e}^{\ii \kvec_{1}\cdot \rvec(x_{1})}
\,{\rm e}^{\ii \kvec_{2}\cdot \rvec(x_{2})}
\ .
}
Viewing again the momenta $\kvec_1$, $\kvec_2$ as charges assigned to the
points $x_1$, $x_2$, the bi-local operator \eExpRep\ corresponds to a dipole,
with charges $\kvec_1=\kvec$, $\kvec_2=-\kvec$, integrated over its internal
charge $\kvec$.
We depict graphically each such dipole as 
\def\legend{
The dipole representing the $\delta$ interaction in \eExpRep.}
\midinsert
\medskip
\centerline{\epsfxsize=3.truecm\epsfbox{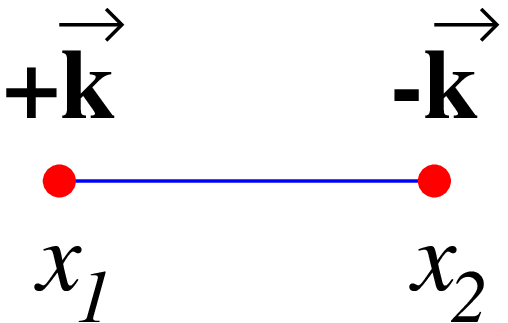}}
\medskip
\noindent
{\bf Fig. 5:\ }
{\sl \legend}
\endinsert
\nfig\fOne{\legend}

Similarly, the product of bi-local operators in \eZPertExp\ and \eCorrFirst\
can be written as an ensemble of $N$ dipoles, that is as the  product of
$2N$ vertex operators with $N$ ``dipolar constraints"
\eqn\eDipConst{
\CC_a\{\kvec_i\}\ =\ (2\pi)^d\delta^d(\kvec_{2a-1}+\kvec_{2a})
\ ,
}
then integrated over all internal charges $\kvec_i$:
\eqn\eProdExp{
\sprod_{a=1}^{N}\delta^{d}(\rvec(x_{2a})-\rvec(x_{2a-1}))=
\int\sprod_{i=1}^{2N}{d^d\kvec_i\over (2\pi)^d}
\sprod_{a=1}^{N}\CC_a\{\kvec_i \}
\sprod_{i=1}^{2N}{\rm e}^{\ii \kvec_i\cdot \rvec(x_i)}.
}
Products of such bi-local operators and of external vertex operators,
as in \eCorrFirst , are depicted by diagrams such as that of Fig. 6.
\def\legend{
Dipole and charges representing bi-local operators and external vertex operators in \eCorrFirst.}
\midinsert
\medskip
\centerline{\epsfxsize=7.5truecm\epsfbox{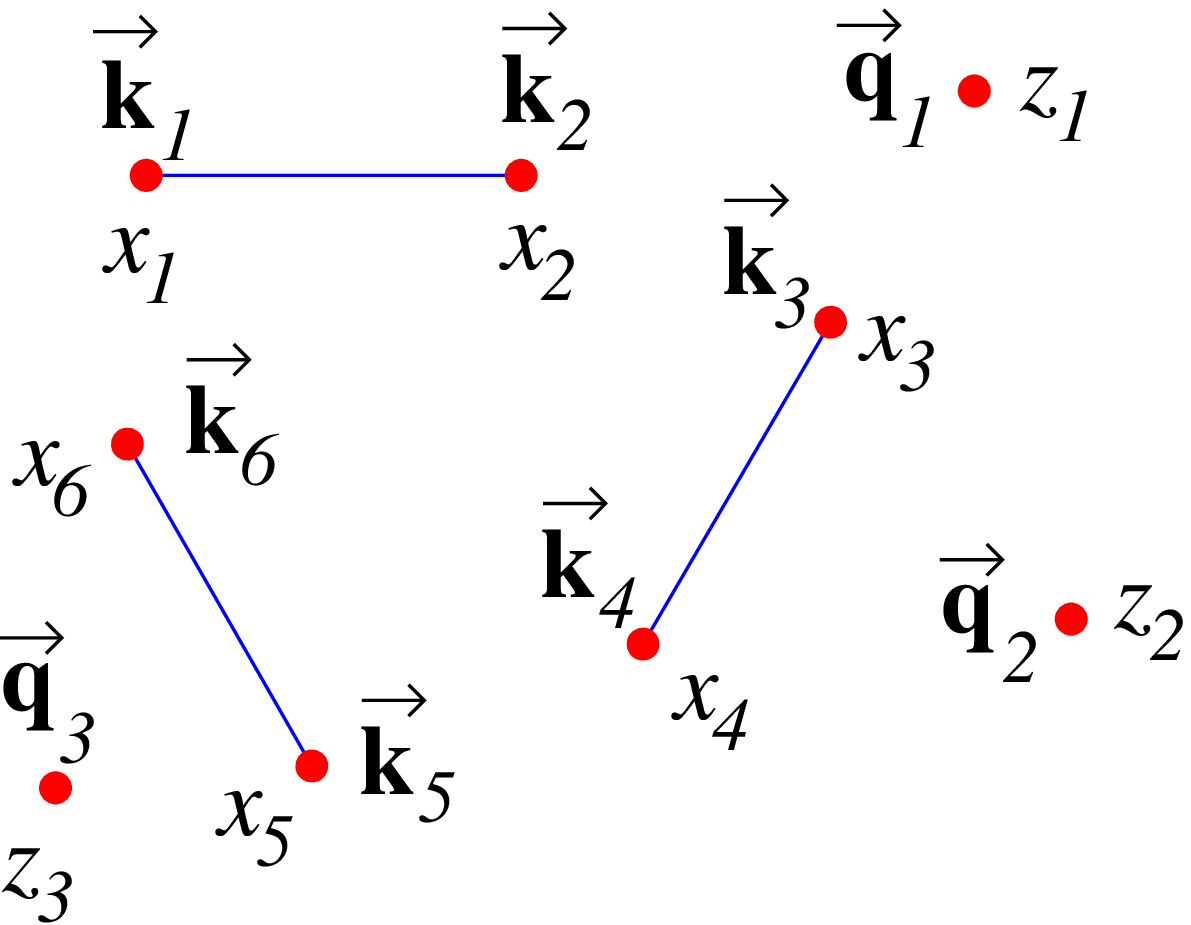}}
\medskip
\noindent
{\bf Fig. 6:\ }
{\sl \legend}
\endinsert
\nfig\fOne{\legend}
The Gaussian average in \eZPertExp, \eProdExp\ is easily performed, and with  the neutrality condition $\sum_i\kvec_i={\vec{\bf 0}}$, we can rewrite it as
\eqn\eGausBis{
\bigl\langle \sprod\limits_{i}{\rm e}^{\ii\kvec_i\cdot\rvec(x_i)}\bigr\rangle_0
\ =\
\exp\Big({-{1\over 2}\sum\limits_{i,j}\kvec_i\cdot\kvec_j
G(x_i-x_j)}
\Big)
\ ,
}
with, as before, the translationally invariant two-point function
\eqn\eMlssProp{
G(x_i-x_j)\ =\
-\,{1\over 2}\,\langle \big({\rvec}(x_i)-{\rvec}(x_j)\big)^2\rangle_0\ =\
-\,{|x_i-x_j|^{2-D}\over (2-D)S_{D}}
\ .
}

Integration over the momenta $\kvec_i$ then gives for the $N$'th
term of the perturbative expansion for the partition function $\CZ$
\eZPertExp\ the ``manifold integral"
\eqn\eZManInt{
Z_N=(2 \pi )^{-Nd/2}\,
\int \sprod_{i=1}^{2N} d^D x_i\ \Delta\{x_i\}^{-{d\over 2}}\, ,
}
with $\Delta\{x_i\}$ the determinant associated with the auxiliary quadratic form
(now on $\RR^{2N}$) $Q\{k_i\}\equiv \ssum\limits_{i,j=1}^{2N} k_i k_j\ G(x_i,x_j)$
restricted to the $N$-dimensional vector space defined by the
$N$ neutrality constraints $\CC_a\{k_i\}$, $k_{2a}+k_{2a-1}=0$.
$\Delta\{x_i\}$ is given explicitly by the determinant of the
$N\times N$ matrix $\Delta_{ab}$
(with row and columns labeled by the dipole indices $a,b=1,\cdots ,N$)
\eqn\eDelExpl{
\Delta\ =\ \det\big(\Delta_{ab}\big),\quad
\Delta_{ab}\ =\
G(x_{2a-1},x_{2b-1})+G(x_{2a},x_{2b})-G(x_{2a-1},x_{2b})-G(x_{2a},x_{2b-1}).
}

Similarly, the $N$'th term in the perturbative expansion of the $P$-point
observable \eCorrFirst\ is
\eqn\eVManInt{
Z_N\left(\{\qvec_l\}\right)=(2\pi)^{-Nd/2}\,
\int \sprod_{i=1}^{2N} d^D x_i\ \Delta\{x_i\}^{-{d\over 2}}\,
\exp\left(-\,{1\over 2}\ssum\limits_{l,m=1}^{P}
\qvec_l\cdot\qvec_m\, {\Delta^{lm}\over\Delta}\right)
.}
$\Delta^{lm}$ is the $(lm)$ minor of the $(P+N)\times(P+N)$
matrix
\eqn\ePplusNMatrix{
\left[
\matrix{
G(z_l,z_m)
&
G(z_l,x_{2b-1})-G(z_l,x_{2b})
\cr
G(x_{2a-1},z_m)-G(x_{2a},z_m)
&
\Delta_{ab}
\cr
}
\right]_{1\le l,m\le P\atop 1\le a,b \le N}
\ .
}

Note that a proper analytic continuation in $D$ of \eZManInt\ and \eVManInt\ 
is insured, as in Section {\bf 2} above, by the use of distance geometry,
where the Euclidean measure over the $x_i$ is understood as
the corresponding measure over the mutual squared distances
$a_{ij}=|x_i-x_j|^2$, a distribution analytic in $D$ \DDGtwo.

\medskip
\subsec{\bf Singular configurations and electrostatics in $\RR^D$}
The integrand in \eZManInt\ is singular  when the determinant vanishes,  
$\Delta\{x_i\}= 0,$  or undefined if the latter becomes negative.
The associated quadratic form $Q\{k_i\}=\ssum\limits_{i,j=1}^{2N} k_i k_j\ G(x_i,x_j)$,
restricted by the $N$ neutrality constraints $\CC_a\{k_i\}$: $k_{2a}+k_{2a-1}=0, a=1,\cdots , N$, is exactly the {\it electrostatic
energy} of a gas of $2N$ scalar charges $k_i$ located at points $x_i$ in $\RR^D$, and constrained to
form $N$ neutral pairs $a$ of charges (dipoles).
For such a globally neutral gas, the Coulomb energy is
{\it minimal} when the charge density is {\it zero everywhere}, {\it i.e.}, when
the non zero charges $k_i$ aggregate into {\it neutral} ``atoms".
When $0<D<2$, because of the vanishing of the Coulomb potential at the origin, $G(0)=0$, the corresponding minimal energy is furthermore {\it zero},
which implies that the quadratic form $Q$ is {\it non-negative}, and thus its determinant is also non-negative:
$\Delta\ge 0$.

Singular $\{x_i\}$ configurations, with $\Delta=0$, still exist when $Q$
is degenerate, which happens when some dipoles are assembled in
such a way that, with appropriate non-zero charges, they still can build
neutral atoms.
This requires some of the points $x_i$ to coincide {\it and} the
corresponding dipoles to form at least one closed loop (Fig. 7).
This ensures that the only sources of divergences are
{\it short-distance singularities}, and extends the Schoenberg theorem
used above.
\def\legend{
The notions of ``atoms'' and ``molecules'', built up from dipoles.}
\midinsert
\medskip
\centerline{\epsfxsize=12.truecm\epsfbox{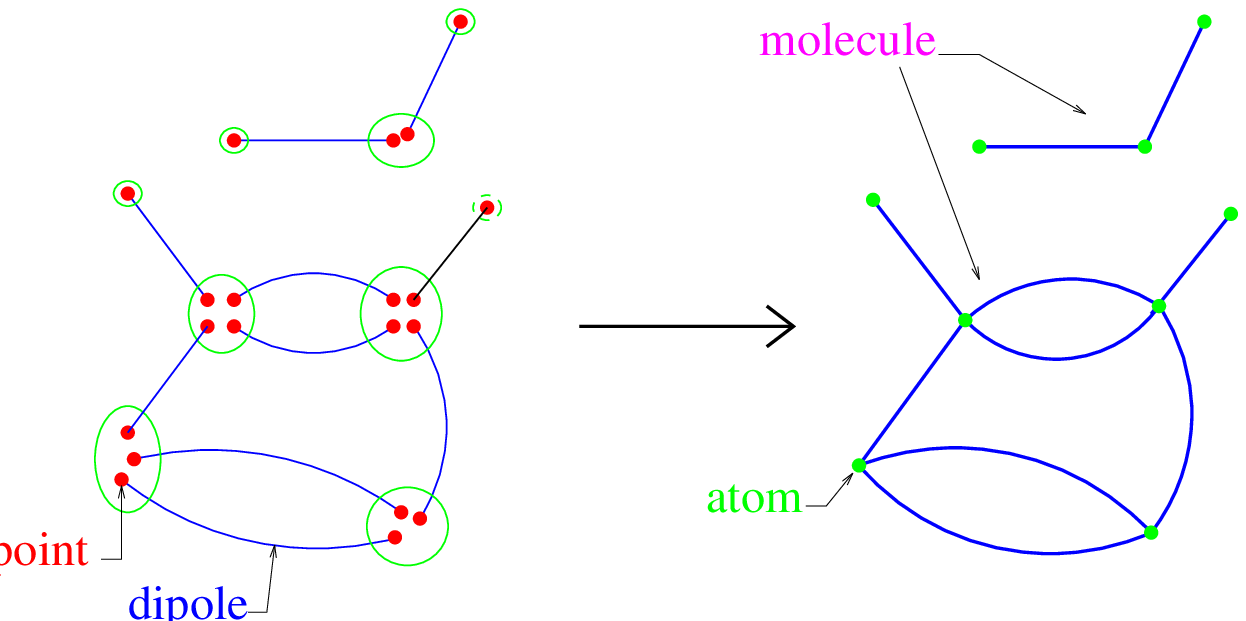}}
\medskip
\noindent
{\bf Fig. 7:\ }
{\sl \legend}
\endinsert
\nfig\fOne{\legend}
\def\legend{
A general diagram with two external points and three internal dipoles, representing bi-local interactions $\delta_a$ (a);
``molecules" describing singular configurations with one (b), two (c,d)
and three (e) ``atoms".
(b,c,d) give UV divergences, (e) does not.}
\midinsert
\medskip
\centerline{\epsfxsize=6.truecm\epsfbox{fig.eps}}
\medskip
\noindent
{\bf Fig. 8:\ }
{\sl \legend}
\endinsert
\nfig\fOne{\legend}
\medskip
\subsec{\bf Multi-local Operator Product Expansion}
A singular configuration can thus be viewed as a connected ``molecule" (Fig. 7), 
characterized by a set $\CM$ of ``atoms" $p$ with assigned positions $x_p$,
and by a set $\CL$ of links $a$ between these atoms, representing the
dipolar constraints $\CC_a$ associated with the $\delta_a \equiv \delta^{d}(\rvec(x_{2a})-\rvec(x_{2a-1}))$ interactions.
For each $p$, we denote by $\CP_p$ the set of charges $i$, at $x_i$, close to point $p$, which
build the atom $p$ and define the  relative (short) distances $y_i=x_i-x_p$ for $i\in \CP_p$ (Fig. 8).

The short-distance singularity of $\Delta^{-d/2}$ is then analyzed by performing a
small $y_i$ expansion of the product of the bilocal operators
$\delta_a$ for the links $a\in \CL$, in the Gaussian manifold theory (Eq. \eCorrFirst).
This expansion around $\CM$ can be written as a
{\it multi-local operator product expansion} (MOPE)
\eqn\eOE{\sprod_{a\in \CL}\delta^{d}(\rvec(x_{2a})-\rvec(x_{2a-1}))=\sum_\Phi \Phi\{x_p\}
\,C^\Phi_{{\underbrace{\scriptstyle \delta\dots\delta}\atop |\CL|}}\{y_i\}}
where the sum runs over all multi-local operators $\Phi$ of the form:
\eqn\eMult{\Phi\{x_p\}=\int d^d\rvec \,
\sprod_{p\in\CM}\Big\{ \,
\npr \left\{\left({\bf \nabla_\rvec}\right)^{q_p}\delta^d(\rvec-\rvec (x_p))
\right\}\,A_p (x_p)\npr\Big\}
}
Here $A_p (x_p)\equiv A^{(r_p,s_p)}\left[\nabla_x\ ;\rvec (x_p)\right]$ is a local operator at
point $x_p$, which is a combination of powers of $x$-derivatives and field $\rvec$,
of degree $s_p$ in $\rvec (x_p)$ and degree $r_p\ge s_p$ in $\nabla_x$.
$({\bf \nabla_\rvec})^{q_p}$ denotes a product of $q_p$ derivatives with
respect to $\rvec$, acting on $\delta^d(\rvec -\rvec (x_p))$. The symbol
``$\npr\quad\npr$" denotes the {\it normal product} subtraction prescription at $x_p$
(which, in a Gaussian average, amounts to setting to zero any derivative of
the propagator ${G}_{ij}$ at coinciding points $x_i=x_j=x_p$).
For ${\rm Card}(\CM)\equiv|\CM|>1$, \eMult\ describes the most general
$|\CM|$-body contact
interaction between the points $x_p$, with possible inserted local operators
$A_p (x_p)$ at each point $x_p$.
For $|\CM|=1$, it reduces to a local operator $A_p(x_p)$.

The coefficient associated with the operator $\Phi$ in the MOPE,
$C^\Phi_{\delta\dots\delta}\{y_i\}$,
can be written as an integral over the momenta $\kvec_i$:
\eqn\eCoeff{ C^{\Phi}_{\delta\dots\delta}\{y_i\} =
\int\kern -.5em
\sprodp_{a\in\CL}
\CC_a\{\kvec_i\}\kern -.2em
\sprod_{p\in\CM}\Bigg\{\sprod_{i\in\CP_p}\kern -.2em d^d\kvec_i\ 
\big\{({\bf \nabla_{\kvec}})^{q_p}
\delta^d(\ssum_{i\in\CP_p}\kvec_i)\big\}\ 
C^{A_p}\{y_i,\kvec_i\}\,
{\rm e}^{-{1\over 2}{\kern -1.5em} \sssum\limits_{\ \quad i,j\in\CP_p}
{\kern -1.5em} \kvec_i.\kvec_j {G}_{ij}}\Bigg\}
}
where $C^{A_p}\{y_i,\kvec_i\}$
is a monomial in the $\{y_i,\kvec_i\}$'s, associated with the operator $A_p$, of similar
global degree $r_p$ in the $\{y_i\}$'s, and $s_p$ in the $\{\kvec_i\}$'s.
The product $\sprodp$ is over all constraints $a\in \CL$ but one.

The MOPE \eOE\  follows from the expression \eExpRep\
in terms of free field exponentials plus constraints, and is established in \DDGfor.

\medskip
\subsec{\bf Power counting and renormalization}
The MOPE \eOE\ allows us to determine those singular configurations which give
rise to actual UV divergences in the manifold integrals \eZManInt\ or \eVManInt.
Indeed, for a given  singular configuration $\CM$,  by integrating
over the domain where the relative positions $y_i=x_i-x_p$ are of order
$|y_i|\ \lesssim\ \rho$, we can use the MOPE
\eOE\ to obtain an expansion of the integrand in \eCorrFirst\
in powers of $\rho$.
Each coefficient $C^\Phi_{\delta\dots\delta}$
gives a contribution of order
$\rho^{\omega_\Phi}$, with degree $\omega_\Phi$ given by power counting as
\eqn\eDegree{\omega_\Phi=D\{2|\CL|-|\CM|\}+d\nu_0\{|\CM|-|\CL|-1\}+
\ssum_{p\in \CM}\big\{\nu_0 (q_p-s_p)+r_p\big\}}
with $\nu_0=(2-D)/2<1$ and $r_p\ge s_p$.
Whenever $\omega_\Phi\le 0$, a UV divergence occurs, as a factor multiplying the
insertion of the corresponding operator $\Phi$.

{\it At the upper critical dimension} $d^\star=2D/\nu_0$, $\omega_\Phi$ becomes
{\it independent} of the number $|\CL|$ of dipoles, and is equal to the canonical
dimension $\omega_\Phi$ of $\int\!\sprod\limits_{p\in \CM}\!d^Dx_p\,\Phi\{x_p\}$
in the Gaussian theory.
\def\legend{
``Molecules'' $\cal M$ producing (a) the one-body local elastic term
$\Phi=\npr (\nabla \rvec_p)^2 \npr$, and (b) the two-body SA interaction term $\Phi
=\delta^d(\rvec_p-\rvec_{p'})$.}
\topinsert
\medskip
\centerline{\epsfxsize=12.truecm\epsfbox{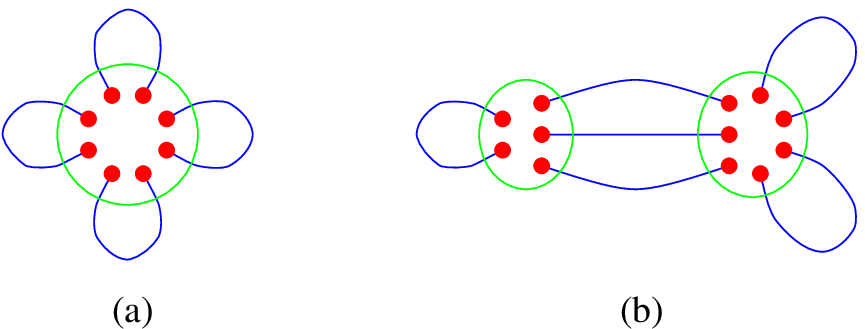}}
\medskip
\noindent
{\bf Fig. 9:\ }
{\sl \legend}
\endinsert
\nfig\fOne{\legend}
Only three relevant multi-local operators $\Phi$, with
$\omega_\Phi\le 0$ and such that the corresponding coefficient does
not vanish by symmetry, are found by simple inspection.
Two of these operators are {\it marginal} ($\omega_\Phi=0$) at  $d^\star$: 
(i) the one-body local elastic term
$\npr (\nabla \rvec_p)^2 \npr$, obtained for $|\CM|=1$ ($q\!=\!0$,
$r\!=\!s\!=2$); 
(ii) the two-body SA interaction term $\delta^d(\rvec_p-\rvec_{p'})$ itself,
obtained through singular configurations with $|\CM|=2$ atoms (and
with $q\!=\!r\!=\!s\!=\!0$ for $p$ and $p'$) (see Fig. 9).

A third operator is {\it relevant} with $\omega_\Phi=-D$, {\it i.e.}, the
identity operator {\bf 1} obtained when $|\CM|=1$ ($q\!=\!r\!=\!s\!=\!0$).
It describes insertions of local ``free energy" divergences along the manifold, proportional to the manifold volume,
which factor out of partition functions like \eZPertExp, to cancel out in correlation functions \eCorrFirst, as already explained 
in part {\bf I}, $\S\, 2.22.$

The above analysis deals with {\it superficial UV divergences} only.
A complete analysis of the general UV singularities associated with
successive contractions toward ``nested" singular configurations
can be performed \DDGfor, using the same techniques as in \DDGtwo\ ({\bf II.} $\S\, 2.8$ above).  
A basic fact is that an
iteration of the MOPE only generates multi-local operators of the same type \eMult .

The results are \DDGfor :
(i) that the observables \eCorrFirst\ are UV finite for $d<d^\star(D)$, and
are meromorphic functions in $d$ with poles at $d=d^\star$;
(ii) that a renormalization operation ${\bf R}$, similar to the subtraction operation
of $\S\ {2.8}$ above \DDGtwo , can be achieved to remove these poles; (iii) that this operation
amounts to a renormalization of the Hamiltonian \eEdwards .

More explicitly, the renormalized correlation functions
$\blangle\sprod\limits_{l=1}^P{\rm e}^{\ii \qvec_l.\rvecR(z_l)}\brangle_{\bf R}$
have a finite perturbative expansion in the renormalized coupling $\bR$,
when $\blangle\cdots\brangle_{\bf R}$ is the average w.r.t the renormalized
Hamiltonian
\eqn\eRenHam{
{\raise.2ex\hbox{$\CH_{\bf R}$}/\raise -.2ex\hbox{${\rm k}_{\rm B}T$}}
\ =\ {1\over 2}\,Z\,\int d^Dx\,\big(\nabla_x\rvecR (x)\big)^2+
{1\over 2}\,{\bR\mu^{\varepsilon/2} Z_b}
\int d^Dx\int d^Dx'\ \delta^{d}\big(\rvecR (x)-\rvecR (x')\big)
\ .}
Here $\mu$ is a renormalization (internal) momentum scale, necessary for infinite manifolds, $\varepsilon=4D-2d\nu_0$;
$Z(\bR)$ and $Z_b(\bR)$ are respectively the field and coupling constant  
renormalization factors, singular at $\varepsilon=0$.

At first order, one finds by explicitly calculating
$C^{(\nabla\rvec)^2}_{\delta}$ and $C^\delta_{\delta\delta}$ that \DDGfor\
$$Z=1+(2\pi A_D)^{-d/2}{\bR \over \varepsilon}{S_D^2(2-D)\over 2D},$$
$$Z_b=1+(2\pi A_D)^{-d/2}{\bR \over \varepsilon}{S_D^2\over 2-D}{\Gamma^2(D/(2-D))\over \Gamma(2D/(2-D))},$$
with $A_D=[S_D(2-D)/2]^{-1}$.
For quantities which do not stay finite in the infinite manifold limit $\CV \to +\infty$, like partition functions, a 
shift in the free energy ({\it i.e.}, an ``additive conterterm'' in $\CH_{\bf R}$), proportional to $\CV$, is also necessary.

Expressing the observables of the SAM model \eEdwards\ in terms of
renormalized variables $\rvec=Z^{1/2}\ \rvecR$, $b=\bR\ \mu^{\varepsilon/2} Z_b\ Z^{d/2}$,
one can derive in the standard way RG equations involving Wilson's functions
$W(\bR)=\raise.2ex\hbox{${\scriptstyle\mu}$}
{\partial\over\partial\mu}\bR\big|_{b}$,
$\nu(\bR)=\nu_0-{1\over 2}\raise.2ex\hbox{${\scriptstyle\mu}$}
{\partial\over\partial\mu}\ln Z\big|_{b}$.
A non-trivial IR fixed point $\bR^\star\!\propto\!\varepsilon$ such that $W(\bR^\star)=0$ is found for
$\varepsilon>0$.
It governs the large distance behavior of the SA infinite manifold, which
obeys scaling laws characterized by the size exponent
$\nu$. The value obtained in this approach,
$\nu=\nu(\bR^\star)$, coincides with that obtained  at first order in $\varepsilon$ in Eq. \NUexp\  above 
[\xref\KN-\xref\BDbis, \xref\DHK].\foot{On can notice the identity between coefficients in the 
renormalization factors  $Z$, $Z_b$ above,  and \aa .}

\medskip
\subsec{\bf Finite size scaling and direct renormalization}
The direct renormalization formalism considered in part {\bf I}, and in {\bf II.} $\S\, 3.2$ above, deals with
{\it finite} manifolds with internal
volume $\CV$, and expresses scaling functions in terms of a dimensionless
second virial coefficient \gsam\ $g=-(2\pi R^2/d)^{-d/2}\CZ_{2,c}/(\CZ_1)^2$,
where $\;\;\;\CZ_1(\CV)(=\CZ/\CZ_0)$ and $\CZ_{2,c}(\CV)$ are respectively the one- and (connected) two-
membrane  partition functions, and $R$ is the effective radius of the membrane.

When dealing with a finite closed manifold (for instance the $D$-dimensional
sphere $\CS_D$ ({\bf II.} $\S\, 2.6$ and \DDGtwo), characterized by its (curved) internal metric,
the massless propagator $G$ gets modified.
Nevertheless, from the short-distance expansion of
$G$ in a general metric,\foot{The expansion at the origin of the massless propagator ${\tilde G}$ on a curved
manifold reads in Riemann normal coordinates
${\tilde G}(x)\simeq G(x)-{|x|^2\over 2D}\blangle\npr(\nabla{\bf r})^2\npr
\brangle$,
with $G(x)\propto |x|^{2-D}$ the propagator in infinite flat space, and
next order terms $\CO(|x|^{4-D})$ proportional to the curvature and
subdominant for $D<2$; the normal product $(\npr\quad\npr)$ is still defined
w.r.t. infinite flat space,
and gives explicitly for a finite manifold with volume $\CV$
$\blangle\npr(\nabla{\bf r})^2\npr\brangle=-1/\CV$.
} 
one can show that
the short-distance MOPE \eOE\ remains valid.  The expansion then extends to multi-local operators $\Phi$ of the form
\eMult, with local operators $A(x)$ which may involve the Riemann curvature
tensor and its derivatives, with appropriate coefficients
$C^\Phi_{\delta\ldots\delta}$ \DDGfor. Still, the coefficients  for those operators $\Phi$
that do not involve derivatives of the metric stay the same as in Euclidean flat space.
  
At $d^\star$, UV divergences still come with insertions of
relevant multi-local operators with $\omega_\Phi\le 0$. When $0<D<2$, the operators involving 
curvature are found by power counting to be {\it all irrelevant}.
Thus the {\it flat infinite membrane} counterterms $Z$ and $Z_b$ still
renormalize the (curved) finite membrane theory. Standard arguments parallel to those of
\BenMa\ 
for polymers then help to establish the direct renormalization formalism (see \DDGfor).
The second virial coefficient $g(b,\CV)$ (as any
{\it dimensionless} scaling function) is UV finite once expressed as a
function $g_{\hbox{\fivebf R}}(\bR, \CV\mu^D)$
of  $\bR$ (and $\mu$).
Then the scaling functions, when expressed in terms
of $g$, obey RG flow equations, and stay {\it finite up to} $\varepsilon=0$.
The existence of a non-trivial IR fixed point $\bR^\star$ for
$\varepsilon>0$ implies that  in the large volume or strong interaction limit, $b\ \CV\ ^{\varepsilon/2D}\to +\infty$, $g$ 
reaches a finite limit $g^\star=g_{\hbox{\fivebf R}}(\bR^\star)$
(independent of $\CV\mu^D$), and so do all scaling functions.
This is just direct renormalization, {\bf QED}.

\medskip
\subsec{\bf Hyperscaling}
Let us first consider a {\it closed} manifold. As mentioned above, the renormalization of partition functions for a
 (finite) SAM requires a shift ($-f_D X^D$) of the free
energy, proportional to the manifold volume $\CV$, and corresponding to the integration of a local 
contact divergence in the bulk. The configuration exponent $\gamma$
is then defined by the scaling of the partition function\foot{Here we consider
$\CZ_1(\CV)\equiv \exp (-f_D  X^D)\ \CZ /\CZ_0,$ {\it i.e.}, the
{\it dimensionally regularized} partition function, which includes the free energy shift (see {\bf I}. $\S\, 2.2.2$).}

\def\sumconf{
\kern -3em \sum_{\quad\qquad\hbox{\sevenrm Configurations}}\kern -3em}
\eqn\eDefGam{\CZ_1(\CV)=\CZ_0^{-1}\int\CD[\rvec]\,
\delta^d(\rvec(0))\, {\rm e}^{-\beta \CH-f_D \CV}\sim \CV^{{\gamma-1\over D}}\ .
}
A consequence of the absence for closed
SAM, for $0<D<2$, of relevant geometrical operators other than the point insertion one, is
 the general hyperscaling law \eHyper\ relating $\gamma$ to $\nu$: $\gamma-1=-\nu d/D.$ 
Indeed, from \eDefGam, $\CZ_1$ is simply multiplicatively renormalized as
$\CZ_1(b,\CV)=Z^{-d/2}\CZ_1^{\bf R}(\bR,\CV\mu^D)$.
This validates the hyperscaling hypothesis that
$\CZ_1\sim \blangle |\rvec|^{-d}\brangle \sim \CV^{-\nu d/D}$.
 Eq. \eHyper\ can been checked explicitly at order $\varepsilon$ for the sphere
$\CS_D$ and the torus $\CT_D$.

For an {\it open} SAM with {\it free} boundaries, and when $1\le D<2$, the
boundary operator
$\int_{\raise-.3ex \hbox{\fiverm boundary}}\kern-2.5em d^{D-1}x\,\hbox{\bf 1}$
is  {\it relevant},  requiring a boundary free energy shift ($-f_{D-1} X^{D-1}$). 
Since  this does not enter into the bulk MOPE, it does not modify the renormalizations of $\rvec$ and $b$.
Furthermore, only for integer $D=1$ is it {\it marginally relevant}  \BDbis, as explained in part {\bf I}; thus
for $D\ne 1$ the hyperscaling relation
\eHyper\ {\it remains valid}.
Only for open polymers at $D=1$, do the corresponding (zero-dimensional) end-point divergences enter the multiplicative
renormalization of $\CZ_1$, and $\gamma$ becomes an independent exponent. In polymer theory, an independent exponent
actually appears for each {\it star vertex}
 \ref\BDJStat{B. Duplantier, {J. Stat. Phys.} {\bf 54} (1989) 581.}.

Previous calculations [\xref\KN,\xref\ArLub,\xref\BDbis] did not involve the massless propagator $\tilde G$
 on a finite manifold with
Neumann boundary conditions, but the simpler propagator $G$ \eMlssProp, corresponding
to a finite SA patch immersed in an infinite Gaussian manifold. The same non-renormalization
argument, as explained in \BDbis\ and in part {\bf I}, yields  $\gamma=1$ for non-integer $D$.

When $D= 2$,  operators involving curvature and boundaries become relevant, and
\eHyper\ is not expected to hold, either for closed or open manifolds.
\medskip
\subsec{\bf $\Theta$-point and long-range interactions}
The above formalism is actually directly applicable to a large class of manifold models
where the interaction can be expressed in terms of free field exponentials
with suitable neutrality constraints $\CC_a\{\kvec_i\}$.
Examples of such interactions are the $n$-body contact potentials, or
 the two-body
long-range Coulomb potential $1/|\rvec-\rvec '|^{d-2}$,
represented by modified dipolar constraints
$\CC\{\kvec_i\}=|\kvec|^{-2}\delta^d(\kvec+\kvec')$.
In these models the MOPE involves the same multi-local operators as in
\eMult , with coefficients \eCoeff\ built with the  corresponding
constraints $\CC_a$.

As an application of the MOPE, one finds that for a polymerized membrane at the $\Theta$-point
where the two-body term $b$ in \eEdwards\ vanishes, the most relevant short-range
interaction is either the usual tricritical {\it three}-body contact potential, with u.c.d.
$d_3^\star=3D/(2-D)$,
as for ordinary polymers \ref\BDter{B. Duplantier, J. Physique {\bf 43} (1982) 991.}, or the two-body singular potential
\hbox{$\Delta_{\rvec}\delta^d(\rvec-\rvec ')$} with u.c.d.
${\tilde d}^\star_2=2(3D-2)/(2-D)$. The latter is the most relevant one when
$D>4/3$ (see \ref\DWthree{K. Wiese and F. David, {Nucl. Phys.} {\bf B 450} (1995) 495, cond-mat/9503126.}).

The very absence of {\it long-range} interactions in the MOPE shows that those
interactions are not renormalized. When considering charged polymerized membranes with a two-body
Coulomb potential for instance, the only (marginally) relevant operator at the upper critical dimension
is the local elastic energy density $\npr (\nabla \rvec )^2 \npr$, which indicates that only
$\rvec$ is renormalized. 
As a consequence, one can show that $\nu=2D/(d-2)$ exactly, generalizing a well-known result for polymers
\ref\dGPf{P. Pfeuty, R. M. Velasco, P.-G. de Gennes, J. Physique {\bf 38} (1977) L5.}.

We did not address here other interesting issues: the approach to the physical $D=2$ case
from the $D<2$ manifold theory \nref\hwa{T. Hwa, Phys. Rev. {\bf A 41} (1990) 1751.}\hskip-.1cm [\xref\DWone,\xref\hwa], 
numerical simulations of 2D polymerized membranes \GK, or
the question of the actual physical phase (crumpled or flat) of a
two-dimensional polymerized membrane in $d$-space \ref\cons{One may consult, {\it e.g.}, K. Wiese,
in {\it Phase Transitions and Critical Phenomena},  C. Domb and J.L. Lebowitz eds.,  vol. 19 (2001),
and the new chapters of the present volume by  M. Bowick, G. Gompper and D.M. Kroll, and L. Radzihovsky.}.  
We have concentrated instead on those more fundamental aspects of renormalization theory, that have been driven by 
the fascinating properties of these fluctuating polymerized membranes.
\bigskip

\bigskip

\centerline{\bf Acknowledgements}
\bigskip

These notes rely heavily on \BD, and on the articles \DDGone, \DDGtwo: {\it Renormalization Theory for
Interacting Crumpled Manifolds}, by Fran\c{c}ois David, B.D., and Emmanuel Guitter. 
For the self-avoiding manifold Edwards model, I followed \DHK, and \DDGthree,   
 \DDGfor:
 {\it Renormalization Theory for Self-Avoiding 
 Polymerized Membranes}, by the same authors. I also wish to thank Emmanuel Guitter for his valuable help with the figures, and 
 Thomas C. Halsey for a careful reading of the manuscript.
\vfill\eject
\listrefs
\bye